# Analysis of five techniques for the internal representation of a digital image inside a quantum processor

Sundaraja Sitharama Iyengar, Latesh K. J. Kumar, and Mario Mastriani
iyengar@cis.fiu.edu, lkumarkj@fiu.edu, mmastria@fiu.edu

*School of Computing & Information Sciences, Florida International University, 11200 S.W. 8th Street, Miami, FL 33199, USA*

In this paper, five techniques, for the representation of a digital image inside a quantum processor, are compared. The techniques are: flexible representation of quantum images (FRQI), novel enhanced quantum representation (NEQR), generalized quantum image representation (GQIR), multi-channel representation for quantum images (MCQI), and quantum Boolean image processing (QBIP). The comparison will be based on implementations on the Quirk simulator, and on the IBM Q Experience processors, from the point of view of performance, robustness (noise immunity), deterioration of the outcomes due to decoherence, and technical viability.

Since its inception nearly two decades ago, Quantum Image Processing (QImP) has always dealt with the same problem, i.e., the internal representation of a digital image inside a quantum circuit efficiently, where such circuits can be optical or of superconductors. In the case of superconducting quantum platforms, they have been freely available to the entire scientific community for approximately five years, which has allowed testing the different techniques for the internal representation of an image on a real physical machine without the need for theoretical speculations. However, during the last five years we have witnessed a complete absence of such implementations. In fact, there are a few examples of them tested on simulators[1,2], e.g., Qiskit of IBM Q Experience[3], leaving the quantum processing units (QPU) from the same company unused, when both options are equally available to the general community. At the same time, other works have appeared, for example[4], where the implementation on QPUs has shown forcefully, and this time on QPUs from different companies[3,5], the problems of various techniques of internal image representation, even on simulators[3,5-8].

From all the accumulated experience in Quantum Information Processing, the scientific community knows that the problem with simulators is that they represent a necessary but not sufficient condition, i.e., if something works in a simulator, e.g. Qiskit, it still needs to be tested on a QPU, but if something does not work in a simulator, then do not even bother to move to the QPU because it is clear that our quantum algorithm under test has problems. In fact, something similar happens with standard versus premium[3] QPUs, where the latter has less decoherence than the former and is generally not as freely accessible as the former, therefore, if we carry out an implementation on a standard QPU with a notable presence of decoherence and the outcomes are not exact, we can associate such inaccuracy with the aforementioned decoherence, but if the implementation is done on a premium QPU with low decoherence and the results are still inaccurate, then the conclusion is obviously, we have problems with our quantum algorithm.

On the other hand, when faced with any technique of internal representation of a digital image inside a QPU, we must ask ourselves two questions:

1. Does this constitute a true Classical-to-Quantum (Cl2Qu) interface?
2. What type of intake does it require?

Before answering the first question, let us briefly analyze the general framework of any scheme of Quantum Image Processing, which is represented in Fig. 1, where:

- Cl2Qu is the internal image representation technique, while
- Qu2Cl means Quantum-to-Classical interface and is related to the quantum measurement technique [9-15] used, which can be weak, strong, or a quantum tomography, all in relation to the Z axis (vertical) of the Bloch's sphere[16-18] of Fig. 2. In particular, this interface requires the orthogonality of the base on which the outcomes are projected, that is, an orthogonal base is necessary for the distinguishability of the outcomes when they are retrieved through the quantum measurement process [9,10,12,14,15].

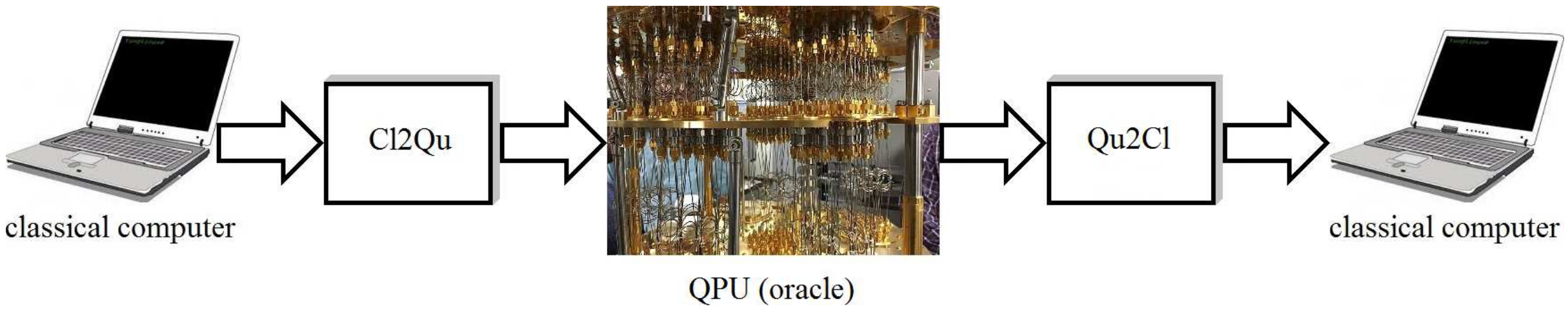

**Figure 1.** Framework of Quantum Image Processing, where Cl2Qu is a classical-to-quantum interface, i.e., the technique for the internal representation of the digital image inside the quantum system (oracle). QPU is the quantum processing unit, and Qu2Cl is the quantum-to-classical interface.

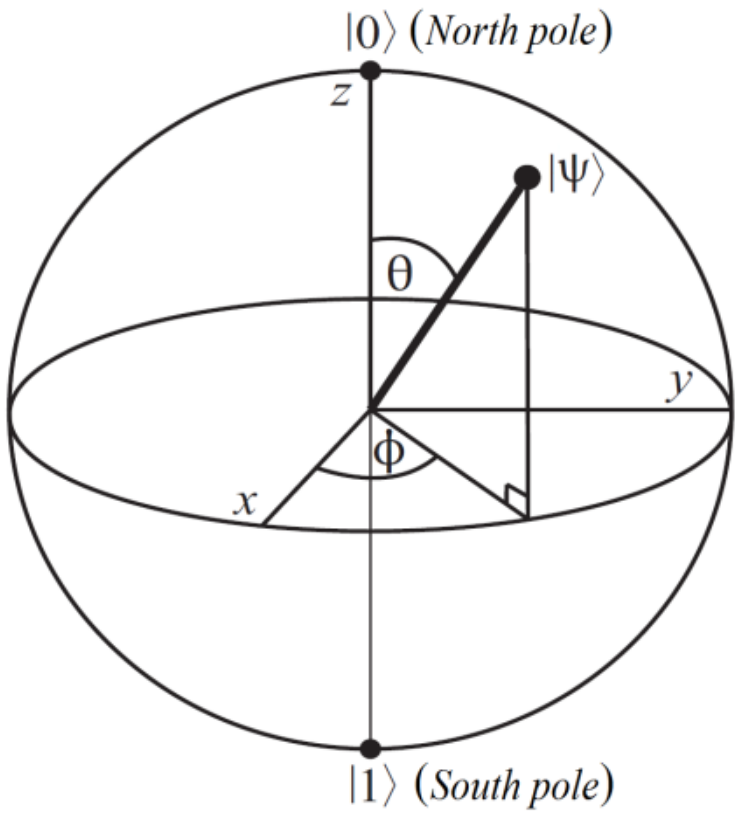

**Figure 2.** Bloch's sphere

On the other hand, Fig. 2 shows the Bloch's sphere, where we can see its two poles:

$$North\ pole = Spin\ up = |0\rangle = \begin{bmatrix} 1 \\ 0 \end{bmatrix}, \text{ and} \tag{1}$$

$$South\ pole = Spin\ down = |1\rangle = \begin{bmatrix} 0 \\ 1 \end{bmatrix}. \tag{2}$$

These poles are called Computational Basis States (CBS). All pure state can be represented on the Bloch's sphere[16-18] of Fig. 2 as a superposition of both CBS $\{|0\rangle, |1\rangle\}$, resulting in a wave-function like the following,

$$|\psi\rangle = \alpha|0\rangle + \beta|1\rangle = \alpha\begin{bmatrix} 1 \\ 0 \end{bmatrix} + \beta\begin{bmatrix} 0 \\ 1 \end{bmatrix} = \begin{bmatrix} \alpha \\ \beta \end{bmatrix} \tag{3}$$

where $|\alpha|^2 + |\beta|^2 = 1$, such that $\alpha \wedge \beta \in \mathbb{C}$ of a Hilbert's space[16]. Specifically, the most practical version of the wave-function in terms of the angles of Fig. 2, will be,

$$|\psi\rangle = cos\frac{\theta}{2}|0\rangle + e^{i\phi}\, sin\frac{\theta}{2}|1\rangle \tag{4}$$

where $0 \le \theta \le \pi$, and $0 \le \phi < 2\pi$.[16] Besides, Equations (3) and (4) are identical for $\alpha = cos\left(\theta/2\right)$ and $\beta = e^{i\phi}\, sin\left(\theta/2\right)$.

As we have previously mentioned, in order to recover the correct outcomes after the Qu2Cl interface, these outcomes must be projected on an orthogonal basis[16] to be distinguishable between them when they are measured[9,10,12,14,15]. If this is not the case, then we are faced with a truly impossible mission given that to distinguish two states with a single measurement there must be orthogonality between these states, a *sine qua non* condition. Then three questions arise:

- Is the distinguishability between orthogonal classical states possible? Yes, it is.
- Is the distinguishability between orthogonal quantum states possible? Yes, it is.
- Is the distinguishability between non-orthogonal quantum states possible? No, it is not. Actually, it is impossible[19].

That is, we must always ask ourselves if the outcomes form an orthogonal base, since quantum measurement loses information from phase to generic qubits and as a consequence of this, infinite outcomes give the same result when in reality they are different.

Complementing all this, the No-Teleportation Theorem[20-22], which is a derivation of the No-Cloning Theorem[16-18], tells us that it is impossible to represent a generic qubit using classical bits, with one exception, i.e., there may be a correspondence biunivocally between qubits and classical bits if the qubit is part of a computational basis state (CBS) $\{|0\rangle,|1\rangle\}$, which represent the North and South poles in the Bloch's sphere of Fig. 2, respectively. Therefore, any technique of internal representation of a digital image inside a QPU has to only deliver CBS to the oracle.

Moreover, the CBS $\{|0\rangle,|1\rangle\}$ are the only exception within Quantum Information Processing[16-18] that:

- remain intact after a quantum measurement[9,10], since they do not lose phase information as in the case of a generic qubit, since for a generic qubit like that of Eq.(3) its probabilities will be:

$$\begin{aligned} Po|0\rangle &= \langle 0|\psi\rangle\langle\psi|0\rangle = \langle 0|(\alpha|0\rangle+\beta|1\rangle)\rangle\langle(\alpha^*\langle 0|+\beta^*\langle 1|)|0\rangle = \\ &= (\alpha\langle 0|0\rangle+\beta\langle 0|1\rangle)(\alpha^*\langle 0|0\rangle+\beta^*\langle 0|1\rangle) = |\alpha|^2, \end{aligned} \tag{5}$$

$$\begin{aligned} Po|1\rangle &= \langle 1|\psi\rangle\langle\psi|1\rangle = \langle 1|(\alpha|0\rangle+\beta|1\rangle)\rangle\langle(\alpha^*\langle 0|+\beta^*\langle 1|)|1\rangle = \\ &= (\alpha\langle 1|0\rangle+\beta\langle 1|1\rangle)(\alpha^*\langle 0|1\rangle+\beta^*\langle 1|1\rangle) = |\beta|^2, \end{aligned} \tag{6}$$

where $(\bullet)^*$ means complex conjugate of $(\bullet)$. While for CBS $\{|0\rangle,|1\rangle\}$, their probabilities will be:

for $|0\rangle$:

$$Po|0\rangle = \langle 0|0\rangle\langle 0|0\rangle = 1, \tag{7}$$

$$Po|1\rangle = \langle 1|0\rangle\langle 0|1\rangle = 0, \tag{8}$$

for $|1\rangle$:

$$Po|0\rangle = \langle 0|1\rangle\langle 1|0\rangle = 0, \tag{7}$$

$$Po|1\rangle = \langle 1|1\rangle\langle 1|1\rangle = 1, \tag{8}$$

that is, the resulting outcomes are the CBSs themselves.

- they can be cloned without problems[23] avoiding the No-Cloning Theorem[16-18], and
- they can be perfectly compared through an *if-then-else* statement[23].

No other type of qubit has these advantages, therefore, any interface between a classic (digital) image and the QPU must deliver CBSs to the latter, since CBSs are the only ones that can be recovered exactly after a quantum measurement in order to faithfully reflect the digital image inside the QPU. Therefore, question one can only be answered analyzing every technique for the internal representation of a digital image inside a QPU in particular.

Regarding question number two, the Quantum Image Processing literature implicitly recognizes two types of intake:

- by wiring, i.e., modifying the quantum circuit according to pixel values of the digital image, and
- by direct input, i.e., intaking data via the input of the oracle itself.

In the first type, we find FRQI[24], NEQR[25], GRQI[26], and MCQI[27], among many others. This group has five serious problems:

(i) *Unnecessary indirect configuration:* In the example of Fig. 3 for FRQI[24], we use the pixel values of the digital image to generate angles with which we modify gates in the quantum circuit. As a result, we obtain the quantum version of those mentioned pixels as outcomes at its output. As we will see in the technique of the second group, we can replace these three steps with a single direct one thanks to an array of CNOT gates[16-18] and without as many gates for each qubit as in the case of FRQI[24], and with the dire consequences it has for the latter, and which we will explain in subsequent items.

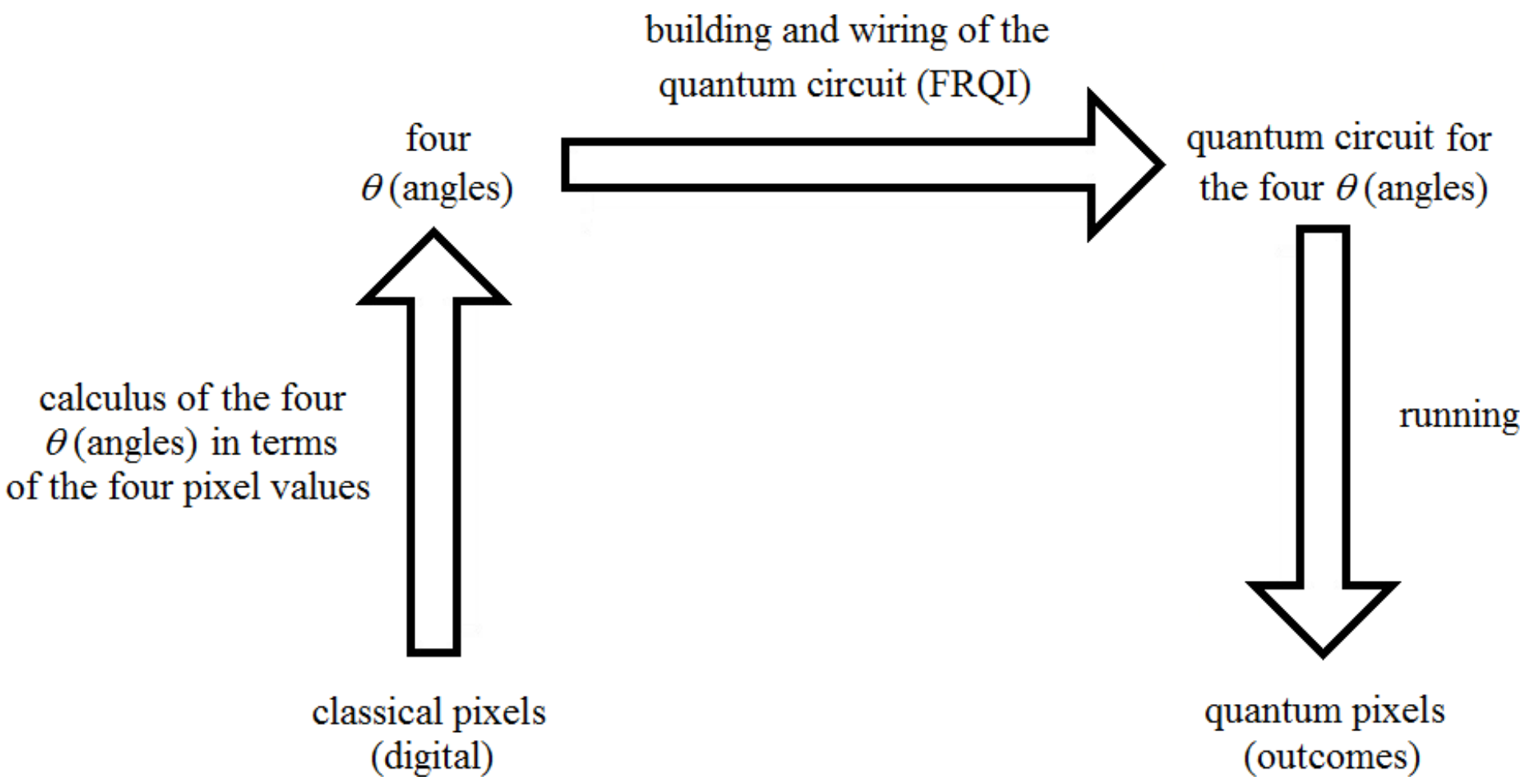

**Figure 3.** FRQI. Passage from the classical pixels of the digital image, grouped by four, to the quantum pixels (outcomes) inside QPU. The passage consists of three steps: Calculation of the four angles based on the values of the four pixels, building and wiring of the quantum circuit, and running of the quantum technique to obtain the outcomes. Note that FRQI does not deliver CBSs to its output.

From Fig. 3, we can see the aggravating factor due to the fact that FRQI[24] does not deliver CBS, with all the dire consequences that this entails regarding the recovery of outcomes via quantum measurement[9,10] and that we will see in the next section.

(ii) *Manual (artisan) modification of the circuit lay-out according to the value of the pixels of the digital image:* In FRQI[24], NEQR[25], GRQI[26], and MCQI[27] the wiring is manual, not automatic, that is, it must be done by a user. If in particular, we refer to the FRQI[24] technique which groups four pixels of the digital image, then for a typical image of 1920 columns, 1080 rows, and 3 channels of color, $1{,}920 \times 1{,}080 \times 3 \div 4 = 1{,}555{,}200$ should be manual accesses to the lay-out of the oracle. It does not take an experimental physicist to understand that this is absurd. Nor is it an excuse that quantum computers are at an early stage in their technological evolution. This does not make practical sense timelessly.

(iii) *The danger of loss of adiabaticity in superconducting QPUs:* Any access to the oracle through a place other than its natural entrances, compromises the adiabaticity of the device *per se*, introducing an excessive and unnecessary additional decoherence[12,13] to the experiment.

(iv) *Complicated concubinage between electronics and optics in an implementation on an optical table*[28]*:* The circuit modification according to the value of the pixels in the case of an optical circuit requires the use of a complex and expensive mixed system made up of electronic and optical devices.

(v) *Accumulation of too many gates per qubit:* It is known from the literature[29] that this inadvisable practice increases the action of decoherence[29].

In the second type, the values of the pixels and their positions are automatically entered (even for color images) via an array of CNOT gates[16-18]. This group is only made up of a single technique: QBIP[30], and as we will see in this work, it is the only one that works correctly in all Quantum Image Processing.

While in the first type, angles are calculated and configurations modified depending on the value of the pixels of the digital image (expressed in bits) so that when executing the technique we obtain the representation of said pixels in their quantum counterpart, which as we explained above must be CBS; QBIP[30], which belongs to the second type, has the following comparative advantages:

- It executes a unique and direct step, Fig. 4, from the digital world to that of the CBS via a simple array of CNOT gates[16-18],
- it does not accumulate gates for each qubit, which is an excellent practice[29],
- it works automatically where the only limitation is the number of qubits of the QPU,
- it has no need to access the quantum circuit other than its natural input,
- it eliminates any possibility of degrading outcomes due to the loss of adiabaticity, and
- it is the one that requires less storage of all the techniques for the representation of a digital image inside a quantum platform.

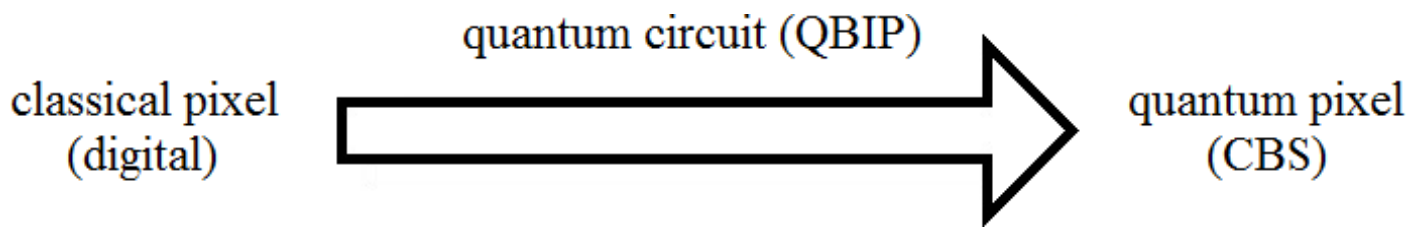

**Figure 4.** QBIP. Passage from the classical pixels of the digital image to the quantum pixels (CBS) inside QPU. The passage only consists of one step.

To summarize, the wiring methods go against everything recommended in superconducting and optical circuits, for this reason the positive statements made in the Quantum Image Processing (QImP) literature[31-47] about these methods are incomprehensible. At the other extreme, we have the three best QImP papers ever written, which among their main merits we can highlight the excellent experimental frameworks carried out in their respective laboratories by the authors[11,48,49], demonstrating that QImP can be done with quality and scientific rigor.

The paper is organized as follows. In Sec. II we analyze the performance of FRQI[24], while Sec. III is reserved for NEQR[25], Sec. IV for GQIR[26], Sec. V for MCQI[27], and Sec. VI for QBIP[30]. Finally, the conclusions are presented in Sec. VII.

## Flexible representation of quantum images (FRQI)

**Theoretical foundations.** FRQI[24] represents a generic image in the following way,

$$\left|I(\theta)\right\rangle = \frac{1}{2^n}\sum_{i=0}^{2^{2n-1}}\left(\cos\theta_i\left|0\right\rangle + \sin\theta_i\left|1\right\rangle\right)\otimes\left|i\right\rangle, \tag{9}$$

$$\theta_i \in \left[0,\frac{\pi}{2}\right],\quad i = 0,1,\ldots,2^{2n}-1, \tag{10}$$

where the angles $\theta_i$ encode the colors of the pixels of the corresponding tile, while $\left|i\right\rangle$ represents the location of the elements of said tile by a horizontal rafter. The FRQI[24] state is a normalized state, i.e., $\left\|I(\theta)\right\| = 1$ as given by

$$\left\|I(\theta)\right\| = \frac{1}{2^n}\sqrt{\sum_{i=0}^{2^{2n}-1}\left(\cos^2\theta_i + \sin^2\theta_i\right)} = 1 \tag{11}$$

Figure 5 shows an example of an image of 2-by-2 pixels and their corresponding codification. FRQI attempts to encode the colors of a classical image through angles. This is done from identity and Hadamard's matrices (in fact, poorly defined on page 66 of the original paper[24]) which are applied on a battery of ancillas of type $\left|0\right\rangle$ and the intervention of the location of the pixels $\left|i\right\rangle$ to arrive at a state $\left|H\right\rangle$. From here, when we apply the matrices of rotation on this state, finally $\left|I(\theta)\right\rangle$ is reached.

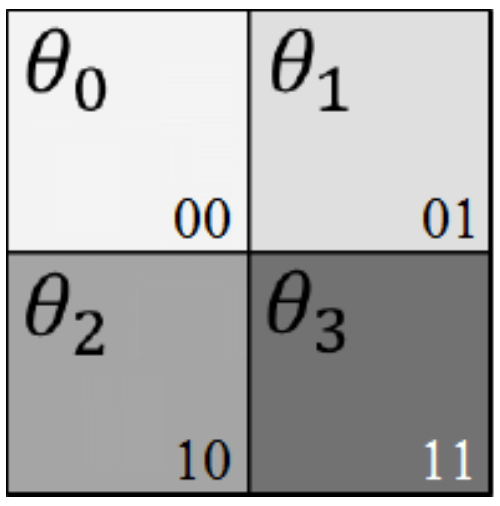

$$|I(1)\rangle = \frac{1}{2}[(\cos\theta_0\,|0\rangle + \sin\theta_0\,|1\rangle) \otimes |00\rangle + (\cos\theta_1\,|0\rangle + \sin\theta_1\,|1\rangle) \otimes |01\rangle + (\cos\theta_2\,|0\rangle + \sin\theta_2\,|1\rangle) \otimes |10\rangle + (\cos\theta_3\,|0\rangle + \sin\theta_3\,|1\rangle) \otimes |11\rangle]$$

**Figure 5.** FRQI for an image of 2-by-2 pixel.

**Implementation on IBM Q[3] Experience.** In this section, we will carry out the implementation of this technique exclusively on IBM Q[3], since Quirk[8] does not provide us with probabilistic outcomes in order to evaluate FRQI in all its magnitude. In this opportunity, we will resort to the implementation of FRQI thanks to the version of Harding and Geetey[2], which we will use in both, a simulator and a physical superconductor machine (QPU), the latter being the first time in the literature of Quantum Image Processing that does such a thing, although we must remember that the authors of this version only tested it on Qiskit and not on an IBM Q[3] QPU. We can see this implementation in Fig. 6.

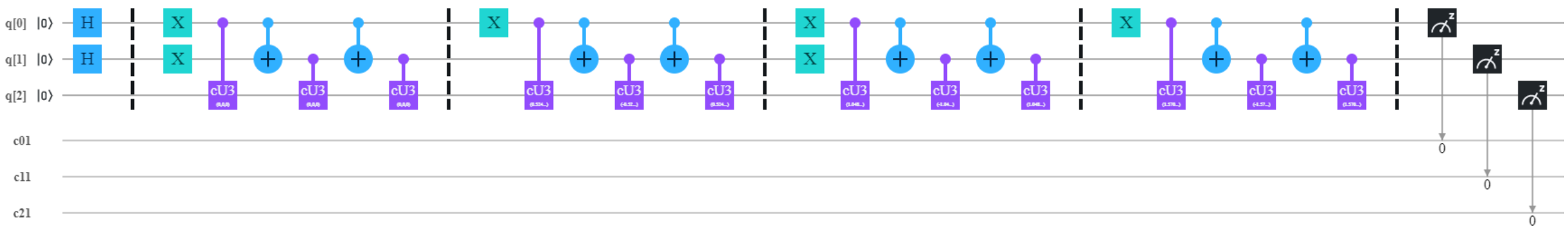

**Figure 6.** Quantum circuit on IBM Q[3] for the Harding and Geetey's version[2] of FRQI.

Based on Figs. 5 and 6, for the following pixel $p_{rc}$ values ($p_{00}$ = 0, $p_{01}$ = 85, $p_{10}$ = 170, $p_{11}$ = 255), which are equivalent to the angles ($\theta_{00}$ = 0, $\theta_{01}$ = π/6, $\theta_{10}$ = π/3, $\theta_{11}$ = π/2), we will obtain the outcomes of Fig. 7, whose statevector is [ 0.5+0j, 0.433+0j, 0.249+0j, 0+0j, 0+0j, 0.25+0j, 0.433+0j, 0.5+0j ].

Beyond the fact that it does not make sense to use a wiring technique for the reasons explained in Section I, the results on the simulator tell us about different types of outcomes which will be difficult to work with in any image treatment process due to the fact that they are not CBS type states. Since as we explained in the previous section, these types of outcomes cannot be cloned, compared, or copied, which seriously conditions the quantum algorithm that is coupled to the output of the FRQI.

The correct coupling scheme between the pixels, their positions, the angles, and the outcomes can be seen in Table I. To understand this coupling correctly, we must look at Fig. 7(a) from the center and towards its two lateral sides, or place a swapping between the outcomes in 000 and 011, and in 001 and 010, and in this way we could see a new version of Fig. 7(a) in the opposite direction, that is, from the ends towards the center. In any case, both Fig 7(a) and 7(b) are symmetrical with respect to an imaginary axis that crosses them through the center between the outcomes with positions 011 and 100.

The results of Table I show a correct correspondence between the three sets of values, i.e., pixels, angles, and outcomes, for an implementation on the IBM Q[3] circuit composer, i.e., a kind of simulator.

| row | column | pixels | angles | outcomes |
|---|---|---|---|---|
| 0 | 0 | 0 | 0 | 0 |
| 0 | 1 | 85 | π/6 | 0.249 |
| 1 | 0 | 170 | π/3 | 0.433 |
| 1 | 1 | 255 | π/2 | 0.5 |

**Table I.** Coupling between the pixel values of Fig. 5, their positions (row, column), the angles, and their respective outcomes according to Fig. 7(a) and (b).

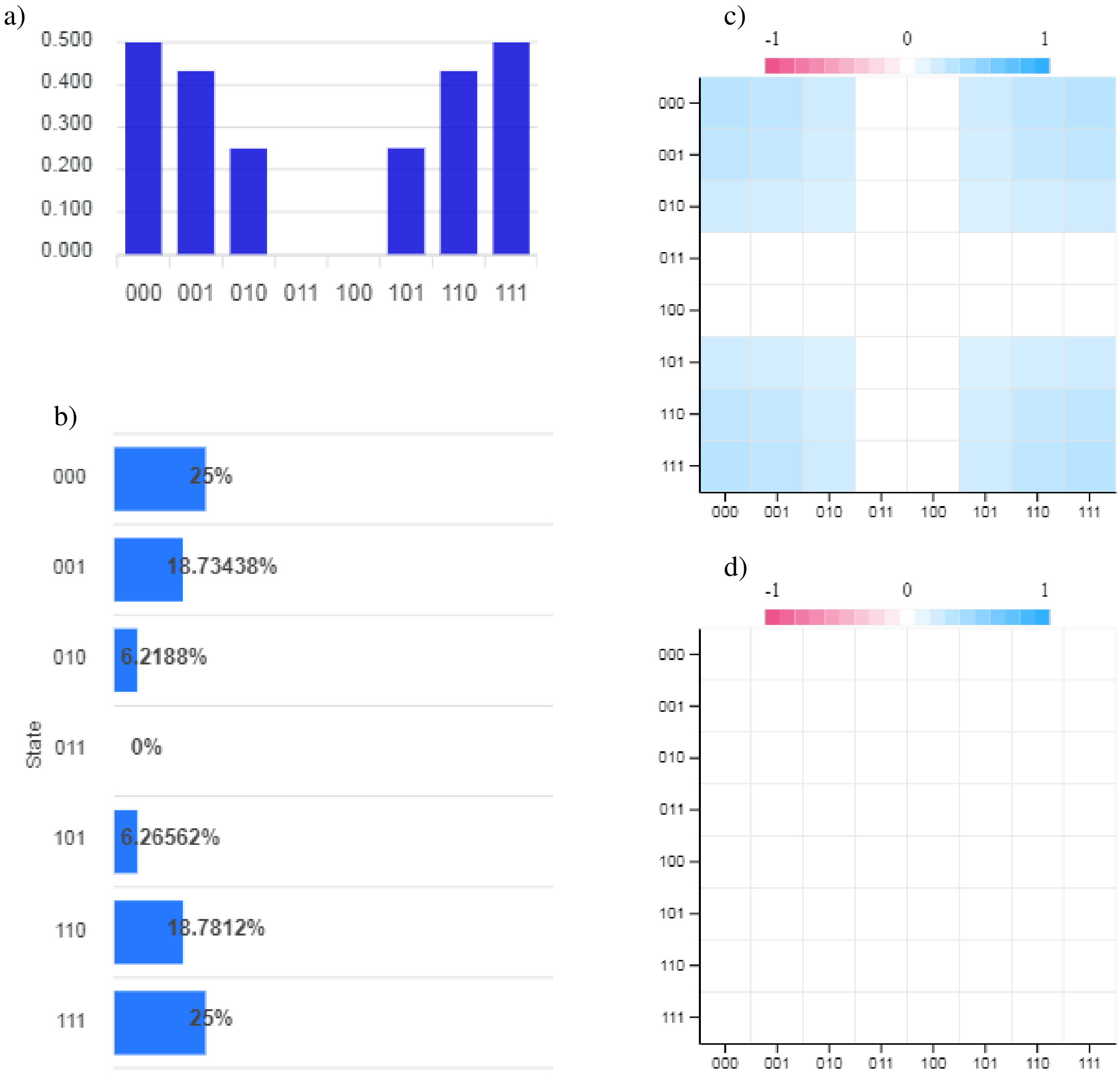

**Figure 7.** Metrics on circuit composer of IBM Q[3] for the quantum circuit of Fig. 6, where the value of the pixels (0, 85, 170, 255) is equivalent to the angles (0, π/6, π/3, π/2), a) the height of the bars is the complex modulus of the wave-function, b) is the real part of the state (histogram) where only those non-zero are shown, c) is the real part density matrix of the state, and d) is the imaginary part density matrix of the state.

In Fig. 9, we follow the same procedure as in Fig. 8 but for the next set of pixel values $p_{rc}$ ($p_{00} = 0$, $p_{01} = 255$, $p_{10} = 0$, $p_{11} = 255$), which are equivalent to the angles ($\theta_{00} = 0$, $\theta_{01} = \pi/2$, $\theta_{10} = 0$, $\theta_{11} = \pi/2$). Then, we obtain the outcomes of Fig. 7, whose statevector is [ 0.5+0j, 0+0j, 0.5+0j, 0+0j, 0+0j, 0.5+0j, 0+0j, 0.5+0j ]. Table II is built with the same criteria that we have established for Table I, in order to analyze the coupling between the pixel values, the angles, and their respective outcomes.

| row | column | pixels | angles | outcomes |
|---|---|---|---|---|
| 0 | 0 | 0 | 0 | 0 |
| 0 | 1 | 255 | π/2 | 0.5 |
| 1 | 0 | 0 | 0 | 0 |
| 1 | 1 | 255 | π/2 | 0.5 |

**Table II.** Coupling between the pixel values of Fig. 5, their positions (row, column), the angles, and their respective outcomes according to Fig. 8(a) and (b).

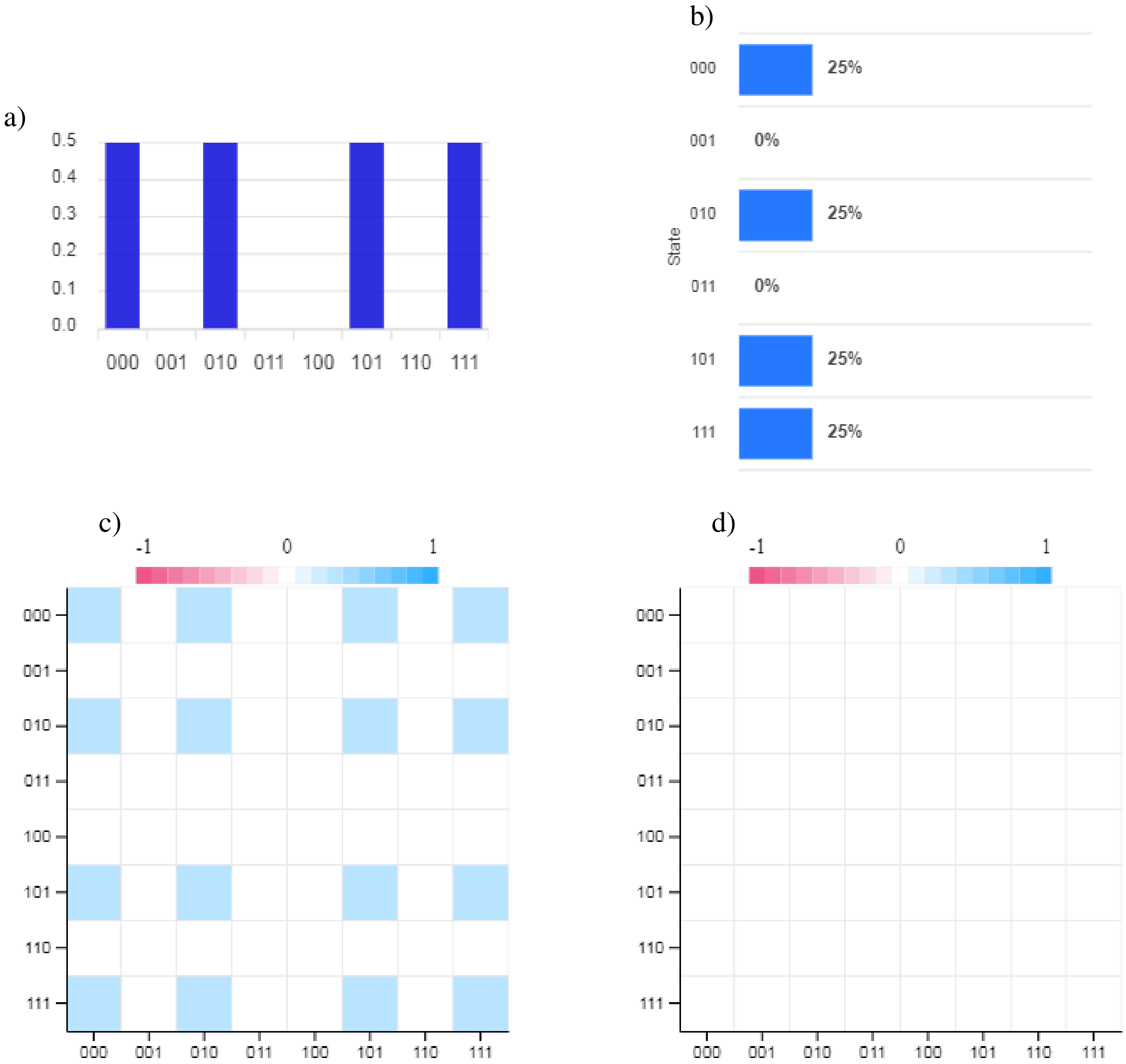

**Figure 8.** Metrics on circuit composer of IBM $Q^3$ for the quantum circuit of Fig. 6, where the value of the pixels (0, 255, 0, 255) is equivalent to the angles (0, π/2, 0, π/2), a) the height of the bars is the complex modulus of the wavefunction, b) is the real part of the state (histogram) where only those non-zero are shown, c) is the real part density matrix of the state, and d) is the imaginary part density matrix of the state.

It is precisely for the same set of pixels in Fig. 8 and Table II ($p_{00} = 0$, $p_{01} = 255$, $p_{10} = 0$, $p_{11} = 255$), which are equivalent to the angles ($\theta_{00} = 0$, $\theta_{01} = \pi/2$, $\theta_{10} = 0$, $\theta_{11} = \pi/2$), that we will carry out the implementation of FRQI on the simulator and a quantum processor (QPU) of IBM $Q^3$ Experience. These implementations can be seen in Fig. 9(a) for the simulator, and in 9(b) for the ibmq_burlington (5 qubits) quantum processor of IBM $Q^3$. As we can see in Fig. 9(a), the IBM $Q^3$ simulator emulates the presence of some decoherence[12,13], while Fig. 9(b) shows us the true result on a QPU, which is obviously very unfavorable, coinciding with identical results from various researchers around the world. This very unfavorable result was explained in a paper[19] in 2017, given that the FRQI outcomes occupy prohibited places in the positions: 001, 011, 100, and 110, which should be zero.

Table III presents a crude comparison of the situation between Figs. 8(b) and 9(b), about what the outcomes for the set of pixels should be and what they actually are. These results should not surprise us in the least, since we must bear in mind that Harding & Geetey[2], from whom we obtained the FRQI protocol in Qiskit, and which required up to 5,000,000,000 shots in Qiskit to achieve their objectives, would be impossible on a QPU, since the greatest number of possible shots is 8192. This shows the amount of effort that FRQI requires to achieve a goal, even when implemented on a simulator.

These results, as well as many others around the world, put a prudent and realistic brake on so many positive affirmations found in the literature[1, 2, 24, 31-37] about the operational benefits of FRQI.

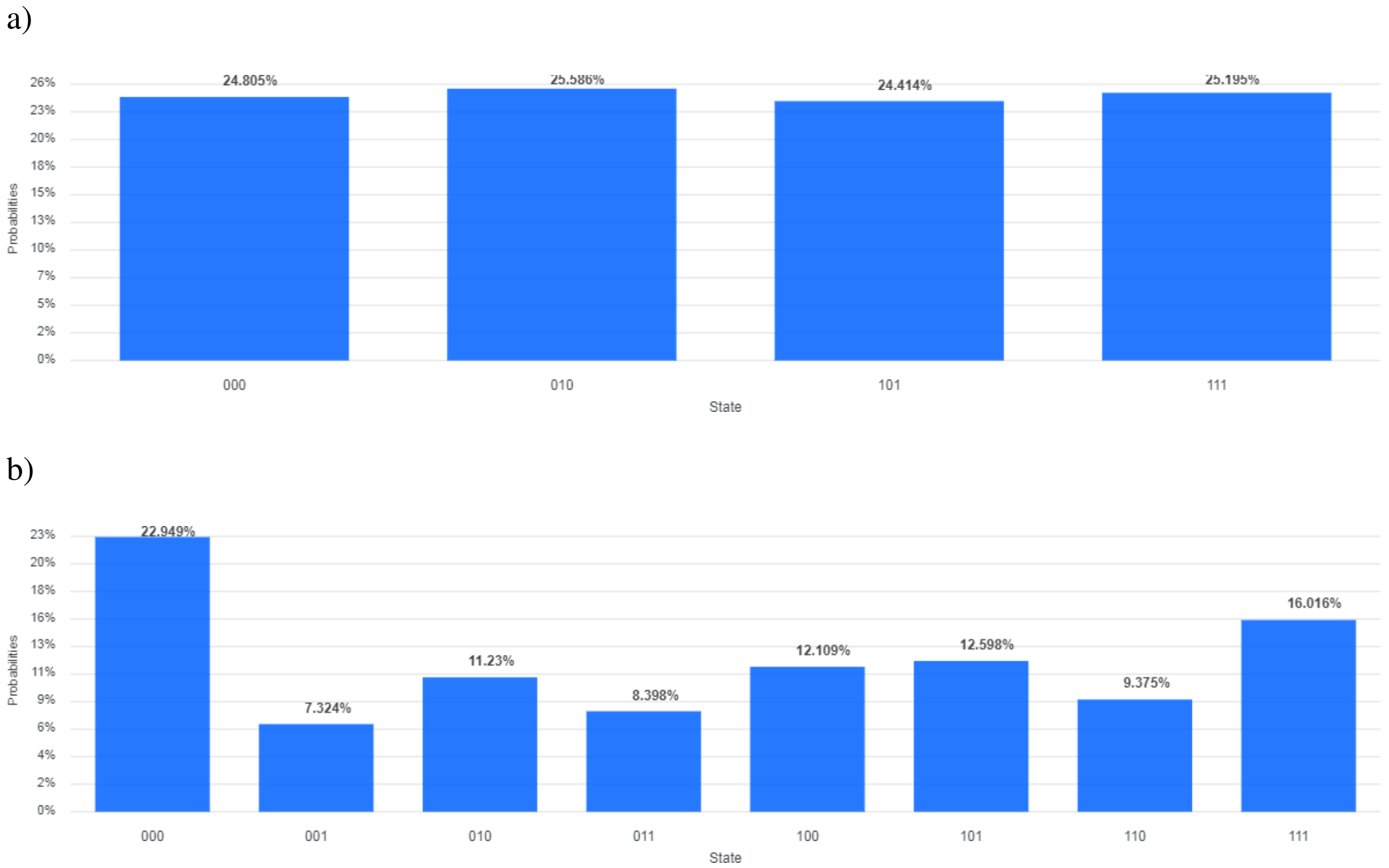

**Figure 9.** Probabilities (outcomes) on IBM Q[3] for the example of Fig. 8, i.e., value of the pixels (0, 255, 0, 255) are equivalent to the angles (0, π/2, 0, π/2): a) on the simulator, b) on the ibmq_burlington (5 qubits) processor, in fairshare Run mode, with 8092 shots.

| row | column | pixels | angles | outcomes (theorists) | outcomes (on the left) | outcomes (on the right) |
|---|---|---|---|---|---|---|
| 0 | 0 | 0 | 0 | 0 | 0.08398 | 0.12109 |
| 0 | 1 | 255 | π/2 | 0.5 | 0.11230 | 0.12598 |
| 1 | 0 | 0 | 0 | 0 | 0.07324 | 0.09375 |
| 1 | 1 | 255 | π/2 | 0.5 | 0.22949 | 0.16016 |

**Table III.** Coupling between the pixel values of Fig. 5, their positions (row, column), the angles, and their respective outcomes according to the theoretical outcomes as well as those of Fig. 9(b).

These results do nothing more than reaffirming what we mentioned in Sec. I about simulators, i.e., if a quantum algorithm has so many difficulties to do its job correctly on a simulator, then, what other results than a total failure can we expect if we try to implement FRQI on a physical platform (QPU). Finally, identical results were reached with the FRQI version of the *citiesatnight*[1] project.

**Novel enhanced quantum representation (NEQR)**

**Theoretical foundations.** To improve FRQI, another technique was proposed in 2013 called NEQR whose model uses two entangled qubit sequences to store the gray-scale and position information; and the whole image in the superposition of the two-qubit sequences. Suppose the gray range of image is $2^q$, binary sequence $C_{YX}^{0}C_{YX}^{1}\ldots C_{YX}^{q-2}C_{YX}^{q-1}$ encodes the gray-scale value $f(Y, X)$ of the corresponding pixel $(Y, X)$ as in:

$$f(Y,X)=C_{YX}^{0}C_{YX}^{1}\ldots C_{YX}^{q-2}C_{YX}^{q-1}, \quad C_{YX}^{k}\in[0,1], \quad f(Y,X)\in\left[0,2^{q}-1\right]. \tag{12}$$

The representative expression of a quantum image for a $2^n \times 2^n$ frame can be written as in:

$$|I\rangle = \frac{1}{2^n}\sum_{Y=0}^{2^{2n-1}}\sum_{X=0}^{2^{2n-1}}|f(Y,X)\rangle|Y\,X\rangle = \frac{1}{2^n}\sum_{Y=0}^{2^{2n-1}}\sum_{X=0}^{2^{2n-1}}\bigotimes_{i=0}^{q-1}|C_{YX}^{i}\rangle|Y\,X\rangle, \tag{13}$$

where, according to their authors[25], both $|C_{YX}^{i}\rangle$ and $|Y\,X\rangle$ are strictly CBS $\{|0\rangle,|1\rangle\}$, otherwise the technique would not work.

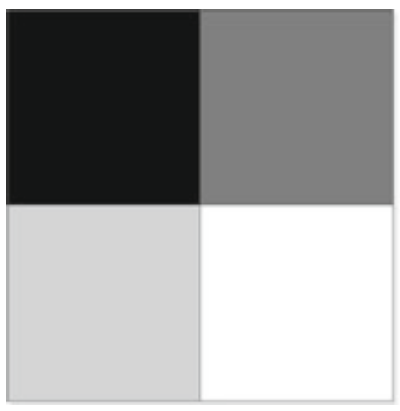

$$|I\rangle = \frac{1}{2}(|0\rangle\otimes|00\rangle + |100\rangle\otimes|01\rangle + |200\rangle\otimes|10\rangle + |255\rangle\otimes|11\rangle)$$
$$= \frac{1}{2}(|00000000\rangle\otimes|00\rangle + |01100100\rangle\otimes|01\rangle + |11001000\rangle\otimes|10\rangle + |11111111\rangle\otimes|11\rangle)$$

**Figure 10.** A 2×2 example image and its representative expression in NEQR.

In Fig. 10, we have a 2×2 NEQR image example. According to this technique[25], making a series of transforms we must go from the representation of a pixel of the digital image of Eq.(12) to its quantum counterpart of Eq.(13). The example in Fig. 10 is represented, in detail, in the following equation,

$$\begin{aligned} |I\rangle &= \frac{1}{2}\begin{pmatrix} |C_{00}^{0}C_{00}^{1}C_{00}^{2}C_{00}^{3}C_{00}^{4}C_{00}^{5}C_{00}^{6}C_{00}^{7}\rangle|00\rangle + |C_{01}^{0}C_{01}^{1}C_{01}^{2}C_{01}^{3}C_{01}^{4}C_{01}^{5}C_{01}^{6}C_{01}^{7}\rangle|01\rangle \\ +|C_{10}^{0}C_{10}^{1}C_{10}^{2}C_{10}^{3}C_{10}^{4}C_{10}^{5}C_{10}^{6}C_{10}^{7}\rangle|10\rangle + |C_{11}^{0}C_{11}^{1}C_{11}^{2}C_{11}^{3}C_{11}^{4}C_{11}^{5}C_{11}^{6}C_{11}^{7}\rangle|11\rangle \end{pmatrix} \\ &= \frac{1}{2}(|00000000\rangle|00\rangle + |01100100\rangle|01\rangle + |11001000\rangle|10\rangle + |11111111\rangle|11\rangle), \end{aligned} \tag{14}$$

where the equivalent image $|I\rangle$ is represented by 4 CBS sequences $C_{YX}^{0}C_{YX}^{1}\ldots C_{YX}^{q-2}C_{YX}^{q-1}$ projected on the respective coordinates also represented by CBS, i.e., $|0\rangle$, and $|1\rangle$. Therefore, we only have to test this technique.

**Implementation on Quirk[8].** We begin with the quantum circuit of NEQR[25] of Fig. 10 on the Quirk platform[8] in Fig. 11, where, the letters *B*, *P* and *D* at the top of the figure mean: *Bloch's sphere*, *probability of* $|1\rangle$, and *density matrix*, respectively. We chose Quirk[8] because it is the platform that best graphically expresses NEQR[25] problems. Moreover, in NEQR, the qubits are prepared according to the quantum transformation of $\Omega_{YX}$ to set a grayscale for the pixel,

$$\Omega_{YX}|0\rangle^{\otimes q} = \bigotimes_{i=0}^{q-1}(\Omega_{YX}^{i}|0\rangle) = \bigotimes_{i=0}^{q-1}|0\oplus C_{YX}^{i}\rangle = \bigotimes_{i=0}^{q-1}|C_{YX}^{i}\rangle = |f(Y,X)\rangle. \tag{32}$$

This group of transformations involves 14 2-*CNOT* gates, 2 Hadamard's gates, and 8 identity matrices for the preparation of one pixel on a grayscale, however, the preparation of the qubits is by hand. From left to right of Fig.11, NEQR consists of 2 steps. Step 1 starts with 10 CBS ground states $|0\rangle$, and 2 Hadamard's gates on the qubits q[8] and q[9] with which the procedure represents the coordinates of the pixel in the tile $(|Y_0\rangle,|X_0\rangle)$.

Now, from Figs. 11 to 14, we implement the example of Fig. 10 when the position of the pixel is $|YX\rangle = |00\rangle$ in Fig. 11, $|YX\rangle = |01\rangle$ in Fig. 12, $|YX\rangle = |10\rangle$ in Fig. 13, and $|YX\rangle = |11\rangle$ in Fig. 14. As we can see in the four figures, $|C_{YX}^{i}\rangle$ are equal for the four positions $|YX\rangle$, which is incorrect, besides none is equal to a CBS, what it should give by definition of the technique, according to Fig. 10 and Eq.(14). In other words, all outcomes are the same for the four positions and different from what they should actually give.

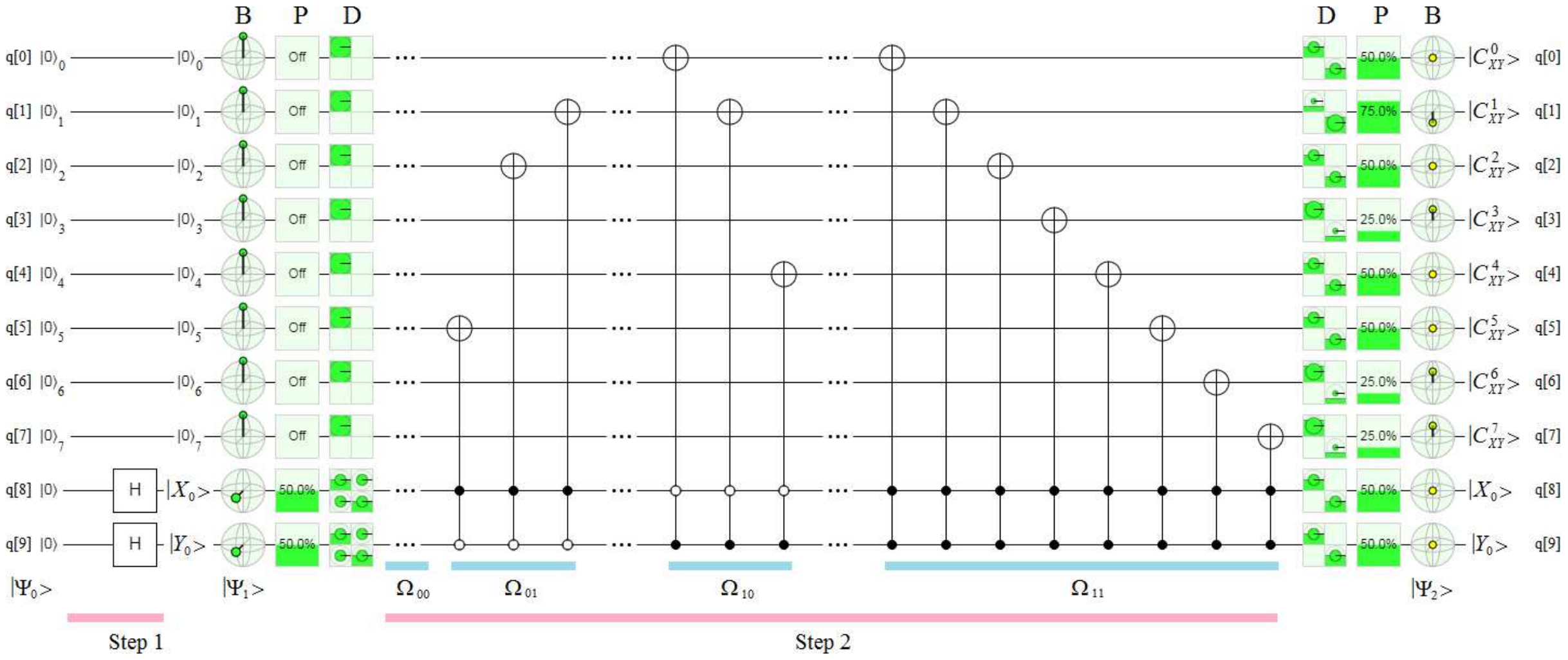

**Figure 11.** Implementation of NEQR on Quirk[8], for example of Fig. 10, $\left|YX\right\rangle = \left|00\right\rangle$.

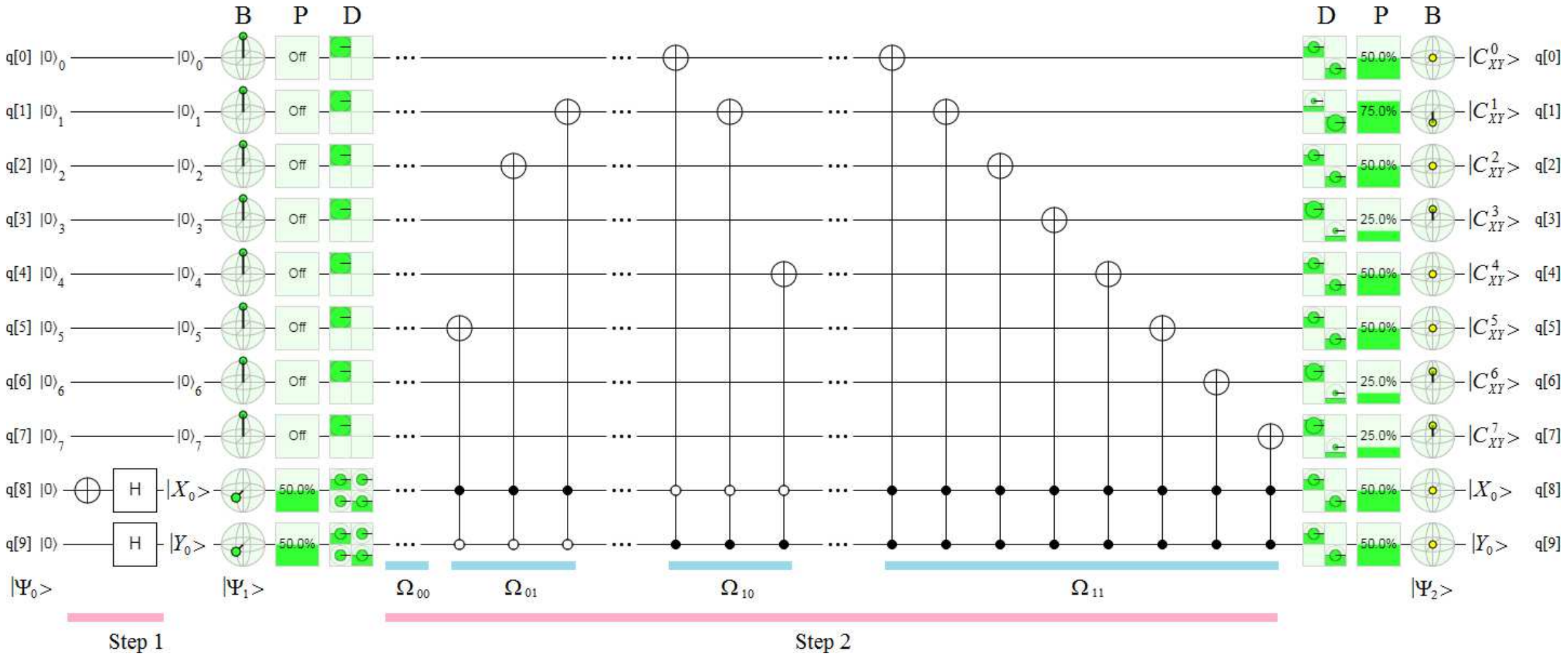

**Figure 12.** Implementation of NEQR on Quirk[8], for example of Fig. 10, $\left|YX\right\rangle = \left|01\right\rangle$.

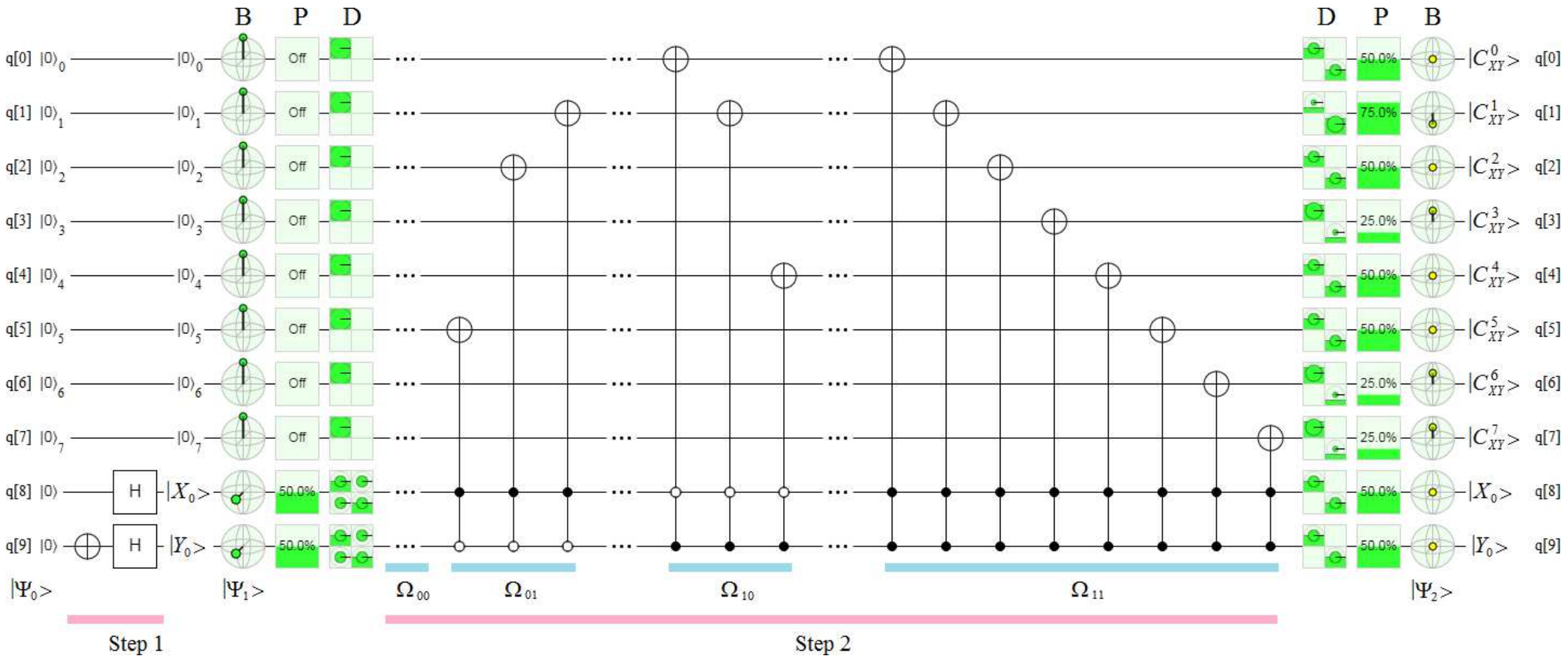

**Figure 13.** Implementation of NEQR on Quirk[8], for example of Fig. 10, $\left|YX\right\rangle = \left|10\right\rangle$.

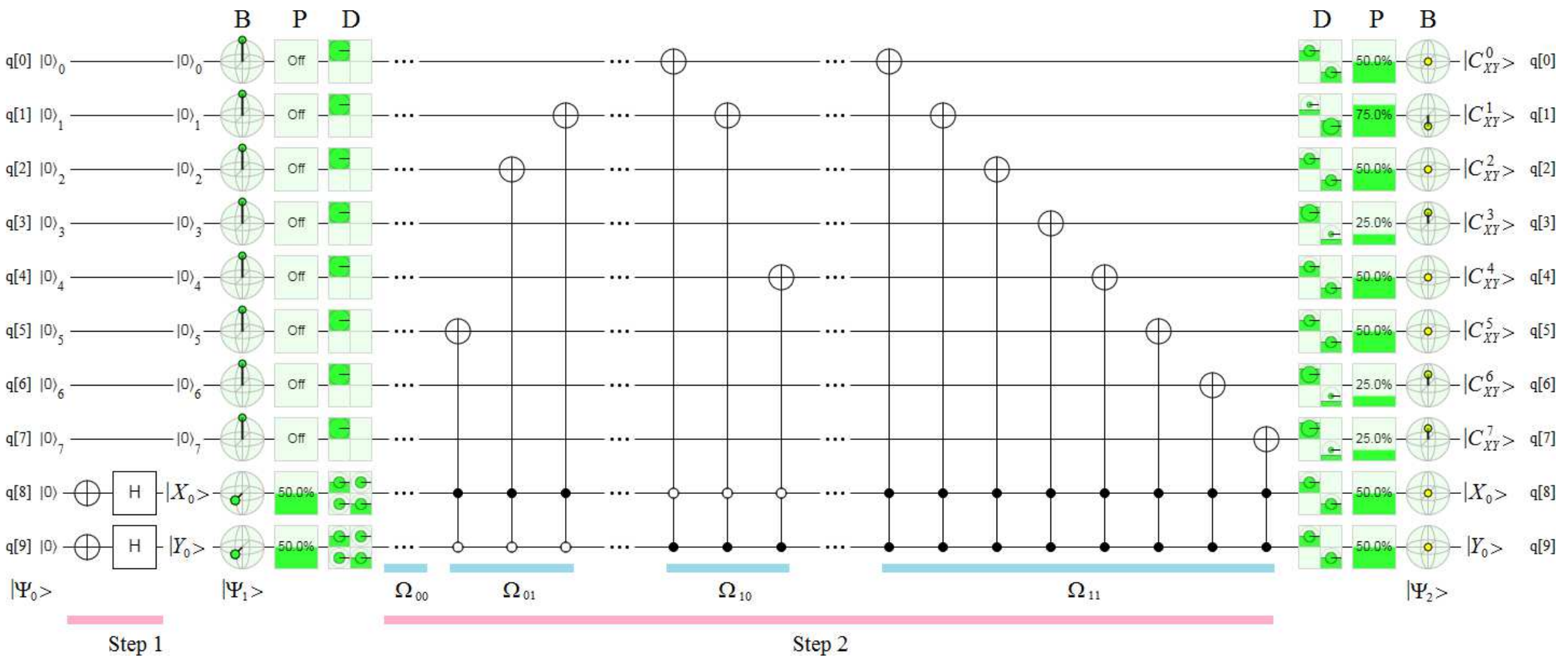

**Figure 14.** Implementation of NEQR on Quirk[8], for example of Fig. 10, $|YX\rangle = |11\rangle$.

Table IV shows the correct outcomes for the NEQR versions of Fig. 10 according to the corresponding positions $|YX\rangle$, while Table V shows the outcomes that the NEQR versions of Fig. 10 really have according to the corresponding positions $|YX\rangle$. The difference is visually evident. The ten qubits were used and none corresponds to a CBS.

| Inputs | | Outputs | | | | | | | | | |
|---|---|---|---|---|---|---|---|---|---|---|---|
| $\|Y\rangle$ | $\|X\rangle$ | $\|Y\rangle$ | $\|X\rangle$ | $\|C_{YX}^7\rangle$ | $\|C_{YX}^6\rangle$ | $\|C_{YX}^5\rangle$ | $\|C_{YX}^4\rangle$ | $\|C_{YX}^3\rangle$ | $\|C_{YX}^2\rangle$ | $\|C_{YX}^1\rangle$ | $\|C_{YX}^0\rangle$ |
| $\|0\rangle$ | $\|0\rangle$ | $\|0\rangle$ | $\|0\rangle$ | $\|0\rangle$ | $\|0\rangle$ | $\|0\rangle$ | $\|0\rangle$ | $\|0\rangle$ | $\|0\rangle$ | $\|0\rangle$ | $\|0\rangle$ |
| $\|0\rangle$ | $\|1\rangle$ | $\|0\rangle$ | $\|1\rangle$ | $\|0\rangle$ | $\|1\rangle$ | $\|1\rangle$ | $\|0\rangle$ | $\|0\rangle$ | $\|1\rangle$ | $\|0\rangle$ | $\|0\rangle$ |
| $\|1\rangle$ | $\|0\rangle$ | $\|1\rangle$ | $\|0\rangle$ | $\|1\rangle$ | $\|1\rangle$ | $\|0\rangle$ | $\|0\rangle$ | $\|1\rangle$ | $\|0\rangle$ | $\|0\rangle$ | $\|0\rangle$ |
| $\|1\rangle$ | $\|1\rangle$ | $\|1\rangle$ | $\|1\rangle$ | $\|1\rangle$ | $\|1\rangle$ | $\|1\rangle$ | $\|1\rangle$ | $\|1\rangle$ | $\|1\rangle$ | $\|1\rangle$ | $\|1\rangle$ |

**Table IV.** Correct outcomes for the NEQR versions of Fig. 10 according to the their corresponding positions $|YX\rangle$.

| Inputs | | Outputs | | | | | | | | | |
|---|---|---|---|---|---|---|---|---|---|---|---|
| $\|Y\rangle$ | $\|X\rangle$ | $\|Y\rangle$ | $\|X\rangle$ | $\|C_{YX}^7\rangle$ | $\|C_{YX}^6\rangle$ | $\|C_{YX}^5\rangle$ | $\|C_{YX}^4\rangle$ | $\|C_{YX}^3\rangle$ | $\|C_{YX}^2\rangle$ | $\|C_{YX}^1\rangle$ | $\|C_{YX}^0\rangle$ |
| $\|0\rangle$ | $\|0\rangle$ | $\|\beta_{00}\rangle$ | $\|\beta_{00}\rangle$ | $\|\beta_{00}\rangle$ | $\|b\rangle$ | $\|\beta_{00}\rangle$ | $\|a\rangle$ | $\|\beta_{00}\rangle$ | $\|\beta_{00}\rangle$ | $\|a\rangle$ | $\|a\rangle$ |
| $\|0\rangle$ | $\|1\rangle$ | $\|\beta_{00}\rangle$ | $\|\beta_{00}\rangle$ | $\|\beta_{00}\rangle$ | $\|b\rangle$ | $\|\beta_{00}\rangle$ | $\|a\rangle$ | $\|\beta_{00}\rangle$ | $\|\beta_{00}\rangle$ | $\|a\rangle$ | $\|a\rangle$ |
| $\|1\rangle$ | $\|0\rangle$ | $\|\beta_{00}\rangle$ | $\|\beta_{00}\rangle$ | $\|\beta_{00}\rangle$ | $\|b\rangle$ | $\|\beta_{00}\rangle$ | $\|a\rangle$ | $\|\beta_{00}\rangle$ | $\|\beta_{00}\rangle$ | $\|a\rangle$ | $\|a\rangle$ |
| $\|1\rangle$ | $\|1\rangle$ | $\|\beta_{00}\rangle$ | $\|\beta_{00}\rangle$ | $\|\beta_{00}\rangle$ | $\|b\rangle$ | $\|\beta_{00}\rangle$ | $\|a\rangle$ | $\|\beta_{00}\rangle$ | $\|\beta_{00}\rangle$ | $\|a\rangle$ | $\|a\rangle$ |

**Table V.** Outcomes that the NEQR versions of Fig. 10 really have according to the corresponding positions $|YX\rangle$.

Respect to Table V, we have:

$|\beta_{00}\rangle = 1/\sqrt{2}\left(|00\rangle + |11\rangle\right)$ with Po|00> = 50%, and Po|11> = 50%, which is a Bell state[16-18], and it is not a CBS $\{|0\rangle, |1\rangle\}$,

$|a\rangle = 0.866|00\rangle + 0.5e^{i\phi_a}|11\rangle$ with Po|00> = 75%, and Po|11> = 25%, i.e., it is not a CBS, and

$|b\rangle = 0.5|00\rangle + 0.866e^{i\phi_b}|11\rangle$ with Po|00> = 25%, and Po|11> = 75%, it is not a CBS too.

Where:

Po|00> means, Probability of |00>,

Po|11> means, Probability of |11>, and

$\phi_a$ and $\phi_b$ are trivial angles in this analysis.

The version that we have used of NEQR corresponds to Fig. 7 of the original paper of its creators[25], in fact, for the same numerical example of its Fig. 3. On the other hand, we have obtained identical results with the alternative version presented by the authors in their Fig. 9. Moreover, we have also obtained identical results with four other quantum platforms, in their simulators as in the case of IBM Q[3], Rigetti[5], Quantum Programming Studio[6], and Quantum Inspire by QuTech[7], as well as their quantum physical machines, as was the case of IBM Q[3] and Rigetti[5]. Therefore, the claims found in the literature[35-37, 39, 47, 50-54] about the benefits of NEQR are incomprehensible, even as a superior version of FRQI, if it does not even work in the most generous of simulators.

**FRI implemented from NEQR.** We chose Quirk[8] again to be able to see the problem in detail that occurs when we want to implement FRQI[24] from NEQR[25]. We resort to the Liu paper[31] (Fig. 1, page 2)

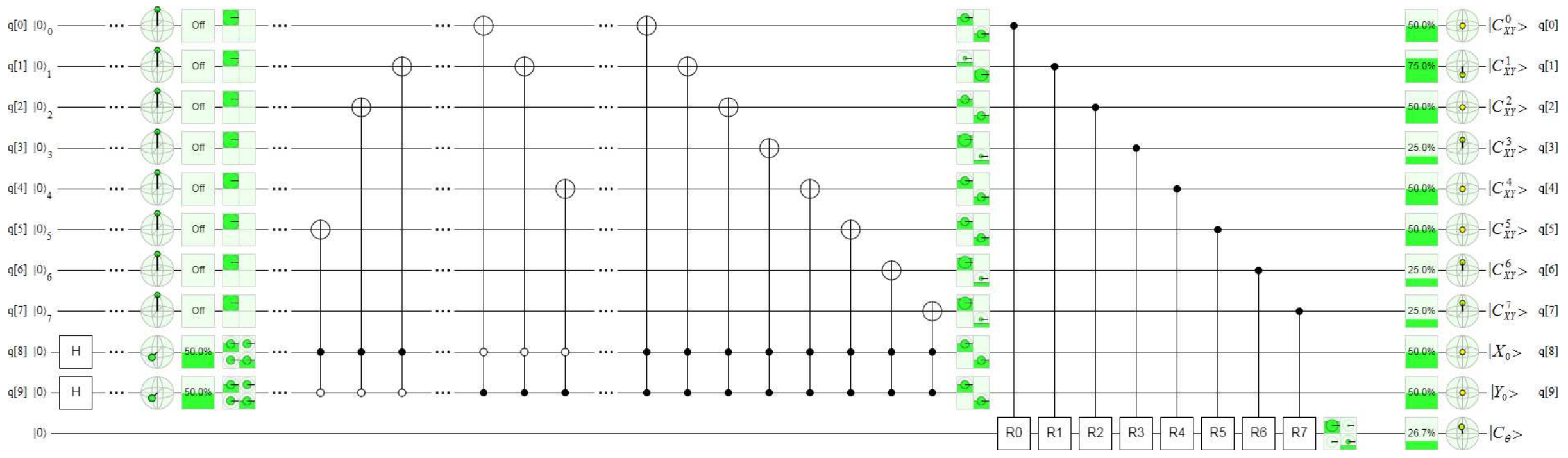

**Figure 15.** Quantum circuit on Quirk[8] for the Liu's (*et al.*) version[31] of FRQI, where the circuit on the right side represents a converter: from NEQR to FRQI, for the first version of NEQR (Fig. 7, page 2843 of its paper[25]).

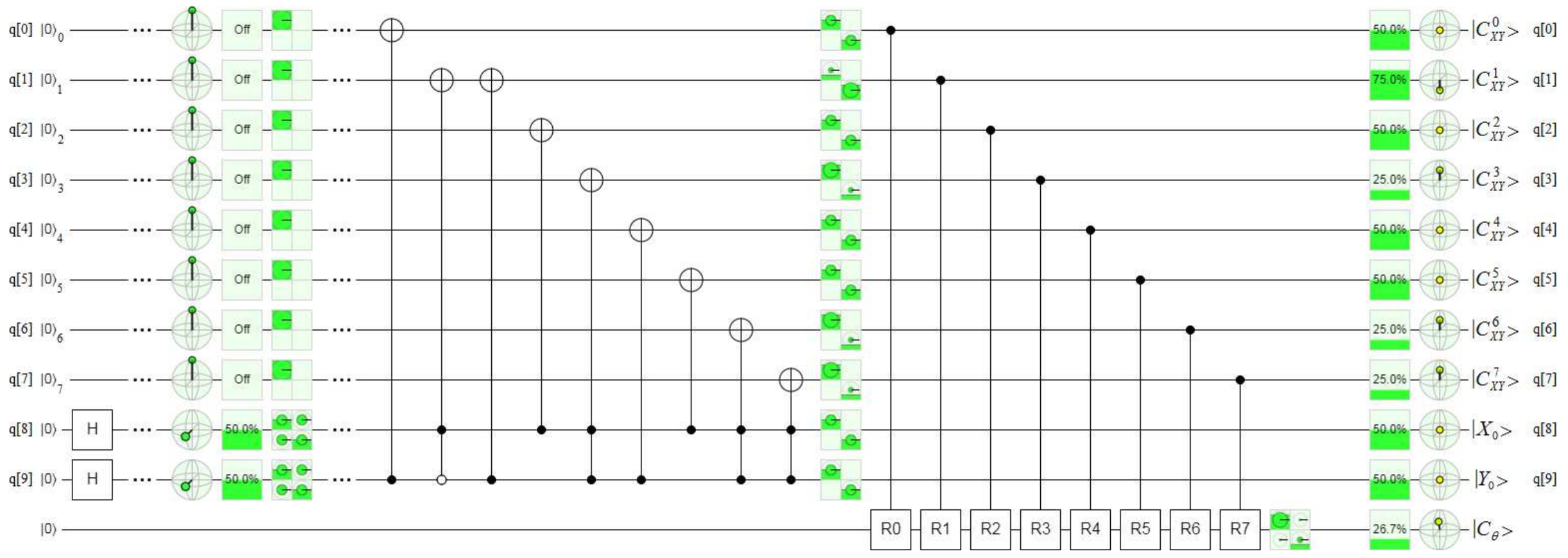

**Figure 16.** Quantum circuit on Quirk[8] for the Liu's (*et al.*) version[31] of FRQI, where the circuit on the right side represents a converter: from NEQR to FRQI, for the second version NEQR (Fig. 9, page 2846 of its paper[25]).

which when applied to the example of Fig. 10 gives us the results of Fig. 15, for the first version of NEQR[25], while in Fig. 16, we obtain the same results for the second version of NEQR[25], i.e., absolutely inconsistent results with the experiment that it supposedly implements. In fact, both outcomes are:

Po|00> = 73.3280 %,
coupling of |0> to <1| = 0.086618,
coupling of |1> to <0| = 0.086618, and
Po|11> = 26.6720 %,

that is, results that have nothing to do with the experiment based on the example in Fig. 10.

To summarize, in this section, we have explored all the options present in the NEQR[35-37, 39, 47, 50-54] literature which shows the same unsatisfactory results.

**Generalized quantum image representation (GQIR)**

**Theoretical foundations.** From[40], we know that GQIR uses $h = \lceil \log_2 H \rceil$ qubits for *Y*-coordinate and $w = \lceil \log_2 W \rceil$ qubits for *X*-coordinate to represent an $H \times W$ image. Both the location information and the color information are captured into normalized quantum states: |0> and |1>. Hence, a GQIR image can be written as

$$|I\rangle = \frac{1}{\sqrt{2}^{h+w}} \left( \sum_{Y=0}^{H-1} \sum_{X=0}^{W-1} |C_{YX}\rangle |YX\rangle \right), \tag{33a}$$

$$|YX\rangle = |Y\rangle |X\rangle = |y_{h-1}\, y_{h-2} \cdots y_0\rangle |x_{w-1}\, x_{w-2} \cdots x_0\rangle,\ y_0, x_0 \in \{0,1\}, \tag{33b}$$

$$|C_{YX}\rangle = |C_{YX}^{q-1} C_{YX}^{q-2} \ldots C_{YX}^{0}\rangle, C_{YX}^{i} \in \{0,1\}, \tag{33c}$$

where, $q$ is the color depth, and

$$h = \begin{cases} \lceil \log_2 H \rceil, H > 1 \\ 1, \qquad H = 1 \end{cases} \tag{34a}$$

$$w = \begin{cases} \lceil \log_2 W \rceil, W > 1 \\ 1, \qquad W = 1 \end{cases} \tag{34b}$$

$|YX\rangle$ is the location information and $|C_{YX}\rangle$ is the color information. It needs $h + w + q$ qubits to represent an $H \times W$ image with gray range $2^q$. Note that GQIR can represent not only gray scale images but also color images because the color depth $q$ is a variable. In most cases, when $q = 2$, the image is binary; when $q = 8$, it is in a gray scale; and when $q = 24$, it is a color image.

According to the Wang *et al.* paper[40], they chose GQIR procedure because:

- It fully exploits the physical properties of qubits (entanglement and superposition) and reduces the number of qubits used to store an image.
- It resolves the real-time computation problem of image processing and provides a flexible method to process any part of a quantum image by using controlled quantum logic gates.
- It can represent a quantum image with any size.
- It is close to the classical image representation. Hence, it is easier for researchers to understand and to transplant a classical image processing algorithm into a quantum system.

Next, we will try to test these claims.

**Implementation on Quirk[8].** We choose Quirk[8] because it is an eminently graphic tool, which will allow us to understand the issue of this technique in all its magnitude. For this purpose, we will resort to two specific examples, one due to Wang's *et al.*[40] and another one extracted from the book by Yan & Venegas-Andraca[35].

We begin with the example of GQIR of Fig. 17, extracted from Fig. 3, page 6 of Wang, *et al.*[40] and that we have implemented in Quirk in Fig. 18 for $|Y_0 X_0\rangle = |00\rangle$.

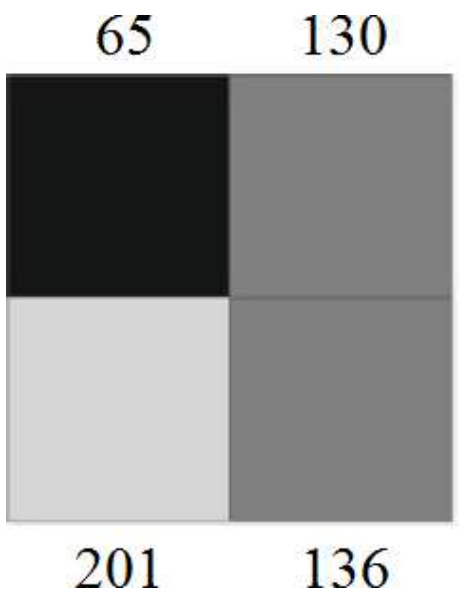

$$|I\rangle = \frac{1}{2}\left(|65\rangle \otimes |00\rangle + |130\rangle \otimes |01\rangle + |201\rangle \otimes |10\rangle + |136\rangle \otimes |11\rangle\right)$$
$$= \frac{1}{2}\left(|01000001\rangle \otimes |00\rangle + |10000010\rangle \otimes |01\rangle + |11001001\rangle \otimes |10\rangle + |10001000\rangle \otimes |11\rangle\right)$$

**Figure 17.** Example of GQIR extracted from Fig. 3, page 6 of Wang's *et al.*[40].

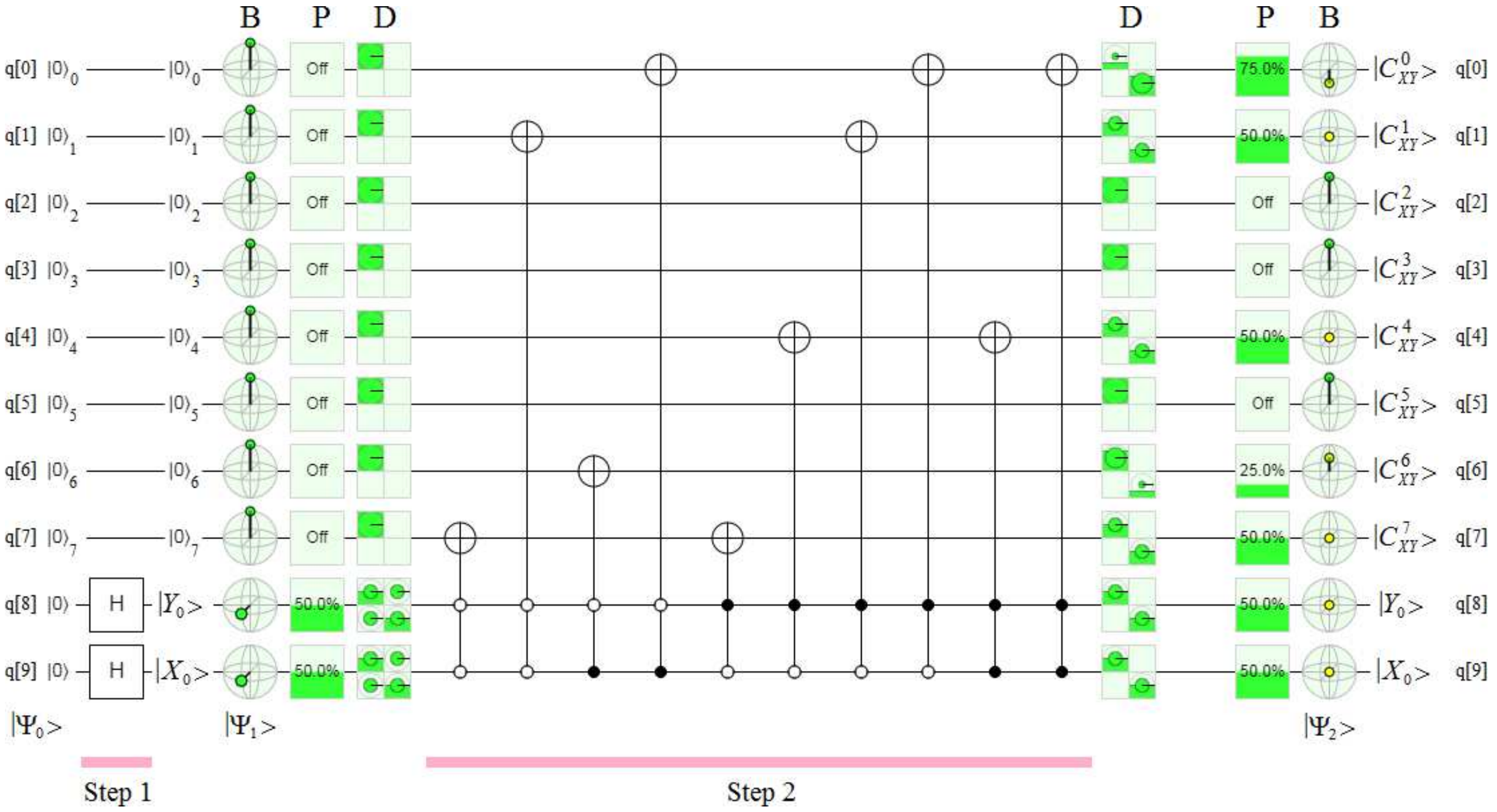

**Figure 18.** Implementation of example of Fig. 17 on Quirk[8].

Going through all possible values of $|Y_0 X_0\rangle$, where, $|Y_0 X_0\rangle \in \{|00\rangle, |01\rangle, |10\rangle, |11\rangle\}$ two tables arise. Table VI shows us the correct outcomes for the GQIR versions of Figs. 17 and 18 in terms of the corresponding inputs $|YX\rangle$, according to Eqs.(33b and 33c). These equations establishes that both $|YX\rangle$ and $|C_{YX}\rangle$ have to be CBS $\{|0\rangle, |1\rangle\}$. However, Table VII shows us the real outcomes as we are giving values to $|YX\rangle$. Both tables coincide in three columns only: 2, 3, and 5. In fact, those columns that were not wired, that is, where GQIR was not applied, because where this technique is applied, it fails. Specifically, values that we saw in Table V for the case of NEQR appear here,

$|\beta_{00}\rangle = 1/\sqrt{2}\left(|00\rangle + |11\rangle\right)$ with Po|00> = 50%, and Po|11> = 50%, i.e., it is not a CBS.

$|a\rangle = 0.866|00\rangle + 0.5e^{i\phi_a}|11\rangle$ with Po|00> = 75%, and Po|11> = 25%, i.e., it is not a CBS.

$|b\rangle = 0.5|00\rangle + 0.866e^{i\phi_b}|11\rangle$ with Po|00> = 25%, and Po|11> = 75%, i.e., it is not a CBS.

$\phi_a$ and $\phi_b$ are trivial in this analysis.

| Inputs | | Outputs | | | | | | | | | |
|---|---|---|---|---|---|---|---|---|---|---|---|
| $\vert Y\rangle$ | $\vert X\rangle$ | $\vert Y\rangle$ | $\vert X\rangle$ | $\vert C_{YX}^{7}\rangle$ | $\vert C_{YX}^{6}\rangle$ | $\vert C_{YX}^{5}\rangle$ | $\vert C_{YX}^{4}\rangle$ | $\vert C_{YX}^{3}\rangle$ | $\vert C_{YX}^{2}\rangle$ | $\vert C_{YX}^{1}\rangle$ | $\vert C_{YX}^{0}\rangle$ |
| $\vert 0\rangle$ | $\vert 0\rangle$ | $\vert 0\rangle$ | $\vert 0\rangle$ | $\vert 1\rangle$ | $\vert 0\rangle$ | $\vert 0\rangle$ | $\vert 0\rangle$ | $\vert 0\rangle$ | $\vert 0\rangle$ | $\vert 1\rangle$ | $\vert 0\rangle$ |
| $\vert 0\rangle$ | $\vert 1\rangle$ | $\vert 0\rangle$ | $\vert 1\rangle$ | $\vert 0\rangle$ | $\vert 1\rangle$ | $\vert 0\rangle$ | $\vert 0\rangle$ | $\vert 0\rangle$ | $\vert 0\rangle$ | $\vert 0\rangle$ | $\vert 1\rangle$ |
| $\vert 1\rangle$ | $\vert 0\rangle$ | $\vert 1\rangle$ | $\vert 0\rangle$ | $\vert 1\rangle$ | $\vert 0\rangle$ | $\vert 0\rangle$ | $\vert 1\rangle$ | $\vert 0\rangle$ | $\vert 0\rangle$ | $\vert 1\rangle$ | $\vert 1\rangle$ |
| $\vert 1\rangle$ | $\vert 1\rangle$ | $\vert 1\rangle$ | $\vert 1\rangle$ | $\vert 0\rangle$ | $\vert 0\rangle$ | $\vert 0\rangle$ | $\vert 1\rangle$ | $\vert 0\rangle$ | $\vert 0\rangle$ | $\vert 0\rangle$ | $\vert 1\rangle$ |

**Table VI.** Correct outcomes for the GQIR versions of Figs. 19 or 20 according to the corresponding inputs $\vert YX\rangle$.

| Inputs | | Outputs | | | | | | | | | |
|---|---|---|---|---|---|---|---|---|---|---|---|
| $\vert Y\rangle$ | $\vert X\rangle$ | $\vert Y\rangle$ | $\vert X\rangle$ | $\vert C_{YX}^{7}\rangle$ | $\vert C_{YX}^{6}\rangle$ | $\vert C_{YX}^{5}\rangle$ | $\vert C_{YX}^{4}\rangle$ | $\vert C_{YX}^{3}\rangle$ | $\vert C_{YX}^{2}\rangle$ | $\vert C_{YX}^{1}\rangle$ | $\vert C_{YX}^{0}\rangle$ |
| $\vert 0\rangle$ | $\vert 0\rangle$ | $\vert \beta_{00}\rangle$ | $\vert \beta_{00}\rangle$ | $\vert \beta_{00}\rangle$ | $\vert a\rangle$ | $\vert 0\rangle$ | $\vert \beta_{00}\rangle$ | $\vert 0\rangle$ | $\vert 0\rangle$ | $\vert \beta_{00}\rangle$ | $\vert b\rangle$ |
| $\vert 0\rangle$ | $\vert 1\rangle$ | $\vert \beta_{00}\rangle$ | $\vert \beta_{00}\rangle$ | $\vert \beta_{00}\rangle$ | $\vert a\rangle$ | $\vert 0\rangle$ | $\vert \beta_{00}\rangle$ | $\vert 0\rangle$ | $\vert 0\rangle$ | $\vert \beta_{00}\rangle$ | $\vert b\rangle$ |
| $\vert 1\rangle$ | $\vert 0\rangle$ | $\vert \beta_{00}\rangle$ | $\vert \beta_{00}\rangle$ | $\vert \beta_{00}\rangle$ | $\vert a\rangle$ | $\vert 0\rangle$ | $\vert \beta_{00}\rangle$ | $\vert 0\rangle$ | $\vert 0\rangle$ | $\vert \beta_{00}\rangle$ | $\vert b\rangle$ |
| $\vert 1\rangle$ | $\vert 1\rangle$ | $\vert \beta_{00}\rangle$ | $\vert \beta_{00}\rangle$ | $\vert \beta_{00}\rangle$ | $\vert a\rangle$ | $\vert 0\rangle$ | $\vert \beta_{00}\rangle$ | $\vert 0\rangle$ | $\vert 0\rangle$ | $\vert \beta_{00}\rangle$ | $\vert b\rangle$ |

**Table VII.** Outcomes that the GQIR versions of Figs. 19 or 20 really have according to the corresponding inputs $\vert YX\rangle$.

The next example was extracted from the Yan and Venegas-Andraca[35] book, its Fig.3.4 (a), of page 60. In fact, Eq.(3.27) of that book says that the outcomes of GQIR must be CBSs $\{\vert 0\rangle,\vert 1\rangle\}$. Figure 19 represents this example, with three pixels different from zero, and five pixels equal to zero.

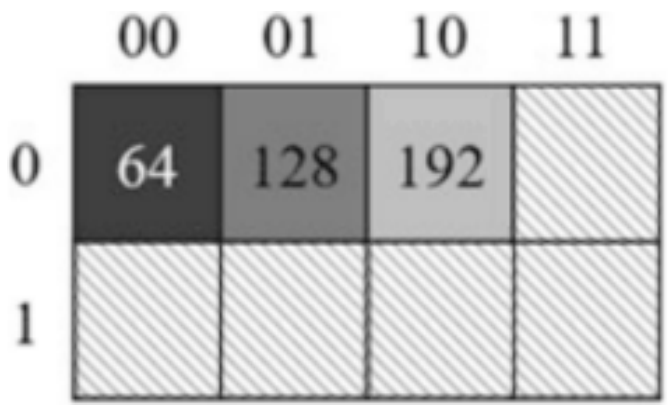

$$\vert I\rangle = \frac{1}{\sqrt{2^3}}\left(\vert 01000000\rangle \otimes \vert 000\rangle + \vert 10000000\rangle \otimes \vert 001\rangle + \vert 11000000\rangle \otimes \vert 010\rangle\right)$$

**Figure 19.** A 1x3 GQIR image and its quantum states, extracted from Fig.3.4, page 60, of Yan and Venegas-Andraca[35].

Figure 20 represents the implementation of Fig. 19, which matches with Fig.3.4 (b) from page 60 of the Yan and Venegas-Andraca[35] book, where, in the case of Fig. 20, the three terms of $\vert I\rangle$ of Fig. 19 are represented with three different colors.

Now, we implement the configuration of Fig. 20 in Quirk[8] in Fig. 21 for $\vert X_1X_0Y_0\rangle = \vert 000\rangle$. With the same criterion as in the previous example, going through all possible values of $\vert X_1X_0Y_0\rangle$, where, $\vert X_1X_0Y_0\rangle \in \{\vert 000\rangle,\vert 001\rangle,\vert 010\rangle,\vert 011\rangle,\vert 100\rangle,\vert 101\rangle,\vert 110\rangle,\vert 111\rangle\}$ two tables arise. Table VIII shows the correct outcomes for the GQIR versions of Figs.19-21 in terms of the corresponding inputs $\vert X_1X_0Y_0\rangle$,

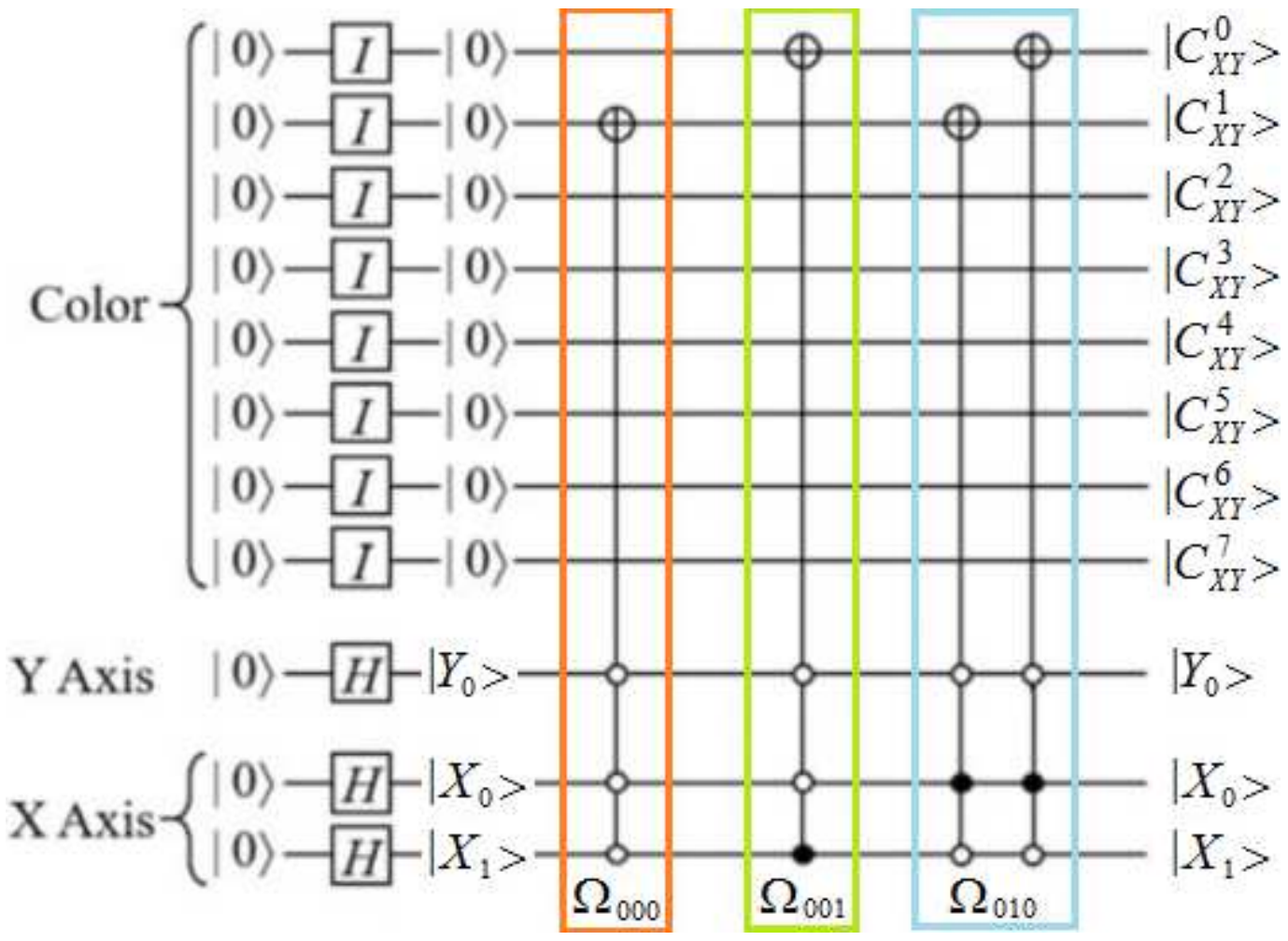

**Figure 20.** Circuit initialization of Fig. 19, extracted from Fig.3.4, page 60, of Yan and Venegas-Andraca[35].

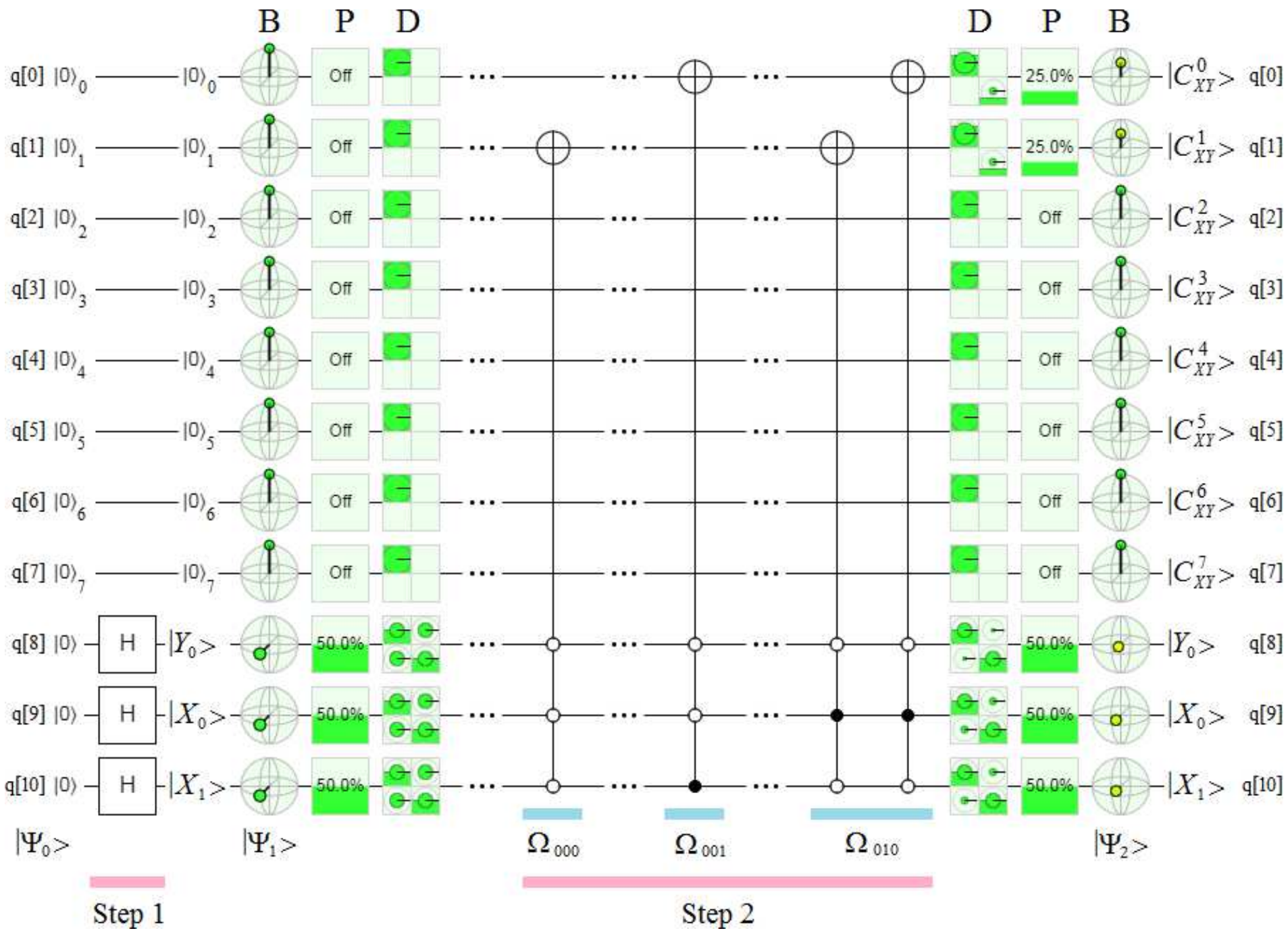

**Figure 21.** Implementation of example of Fig. 19 on Quirk[8].

where here too, both $\left|X_1X_0Y_0\right\rangle$ and $\left|C_{YX}\right\rangle$ have to be CBS $\{|0\rangle,|1\rangle\}$. However, Table IX shows us the real outcomes as we are giving values to $\left|X_1X_0Y_0\right\rangle$. In this case, there is a coincidence in six columns of both tables, from the second one to the seventh one, in fact, those that were not wired, that is, where GQIR was not applied, because where this technique is applied, it fails again. Table IX shows the differences in detail, where:

$|a\rangle = 0.866|00\rangle + 0.5e^{i\phi_a}|11\rangle$, with Po|00> = 75%, and Po|11> = 25%,

$|b\rangle = [0.6959 \quad 0.125 \quad 0.125 \quad 0.6959]^T$, with

Po|00> = 48.4375%, Po|01> = 1.5625%, Po|10> = 1.5625%, and Po|11> = 48.4375%,

$|c\rangle = [0.6614 \quad 0.25 \quad 0.25 \quad 0.6614]^T$, with

Po|00> = 43.75%, Po|01> = 6.25%, Po|10> = 6.25%, and Po|11> = 43.75%,

| Inputs | | | Outputs | | | | | | | | | | |
|---|---|---|---|---|---|---|---|---|---|---|---|---|---|
| $\vert X_1\rangle$ | $\vert X_0\rangle$ | $\vert Y_0\rangle$ | $\vert X_1\rangle$ | $\vert X_0\rangle$ | $\vert Y_0\rangle$ | $\vert C_{YX}^7\rangle$ | $\vert C_{YX}^6\rangle$ | $\vert C_{YX}^5\rangle$ | $\vert C_{YX}^4\rangle$ | $\vert C_{YX}^3\rangle$ | $\vert C_{YX}^2\rangle$ | $\vert C_{YX}^1\rangle$ | $\vert C_{YX}^0\rangle$ |
| $\vert 0\rangle$ | $\vert 0\rangle$ | $\vert 0\rangle$ | $\vert 0\rangle$ | $\vert 0\rangle$ | $\vert 0\rangle$ | $\vert 0\rangle$ | $\vert 0\rangle$ | $\vert 0\rangle$ | $\vert 0\rangle$ | $\vert 0\rangle$ | $\vert 0\rangle$ | $\vert 1\rangle$ | $\vert 0\rangle$ |
| $\vert 0\rangle$ | $\vert 1\rangle$ | $\vert 0\rangle$ | $\vert 0\rangle$ | $\vert 1\rangle$ | $\vert 0\rangle$ | $\vert 0\rangle$ | $\vert 0\rangle$ | $\vert 0\rangle$ | $\vert 0\rangle$ | $\vert 0\rangle$ | $\vert 0\rangle$ | $\vert 0\rangle$ | $\vert 1\rangle$ |
| $\vert 1\rangle$ | $\vert 0\rangle$ | $\vert 0\rangle$ | $\vert 1\rangle$ | $\vert 0\rangle$ | $\vert 0\rangle$ | $\vert 0\rangle$ | $\vert 0\rangle$ | $\vert 0\rangle$ | $\vert 0\rangle$ | $\vert 0\rangle$ | $\vert 0\rangle$ | $\vert 1\rangle$ | $\vert 1\rangle$ |
| $\vert 1\rangle$ | $\vert 1\rangle$ | $\vert 0\rangle$ | $\vert 1\rangle$ | $\vert 1\rangle$ | $\vert 0\rangle$ | $\vert 0\rangle$ | $\vert 0\rangle$ | $\vert 0\rangle$ | $\vert 0\rangle$ | $\vert 0\rangle$ | $\vert 0\rangle$ | $\vert 0\rangle$ | $\vert 0\rangle$ |
| $\vert 0\rangle$ | $\vert 0\rangle$ | $\vert 1\rangle$ | $\vert 0\rangle$ | $\vert 0\rangle$ | $\vert 1\rangle$ | $\vert 0\rangle$ | $\vert 0\rangle$ | $\vert 0\rangle$ | $\vert 0\rangle$ | $\vert 0\rangle$ | $\vert 0\rangle$ | $\vert 0\rangle$ | $\vert 0\rangle$ |
| $\vert 0\rangle$ | $\vert 1\rangle$ | $\vert 1\rangle$ | $\vert 0\rangle$ | $\vert 1\rangle$ | $\vert 1\rangle$ | $\vert 0\rangle$ | $\vert 0\rangle$ | $\vert 0\rangle$ | $\vert 0\rangle$ | $\vert 0\rangle$ | $\vert 0\rangle$ | $\vert 0\rangle$ | $\vert 0\rangle$ |
| $\vert 1\rangle$ | $\vert 0\rangle$ | $\vert 1\rangle$ | $\vert 1\rangle$ | $\vert 0\rangle$ | $\vert 1\rangle$ | $\vert 0\rangle$ | $\vert 0\rangle$ | $\vert 0\rangle$ | $\vert 0\rangle$ | $\vert 0\rangle$ | $\vert 0\rangle$ | $\vert 0\rangle$ | $\vert 0\rangle$ |
| $\vert 1\rangle$ | $\vert 1\rangle$ | $\vert 1\rangle$ | $\vert 1\rangle$ | $\vert 1\rangle$ | $\vert 1\rangle$ | $\vert 0\rangle$ | $\vert 0\rangle$ | $\vert 0\rangle$ | $\vert 0\rangle$ | $\vert 0\rangle$ | $\vert 0\rangle$ | $\vert 0\rangle$ | $\vert 0\rangle$ |

**Table VIII.** Correct outcomes for the GQIR versions of Figs. 20 or 21 according to the corresponding inputs $\vert X_1X_0Y_0\rangle$.

| Inputs | | | Outputs | | | | | | | | | | |
|---|---|---|---|---|---|---|---|---|---|---|---|---|---|
| $\vert X_1\rangle$ | $\vert X_0\rangle$ | $\vert Y_0\rangle$ | $\vert X_1\rangle$ | $\vert X_0\rangle$ | $\vert Y_0\rangle$ | $\vert C_{YX}^7\rangle$ | $\vert C_{YX}^6\rangle$ | $\vert C_{YX}^5\rangle$ | $\vert C_{YX}^4\rangle$ | $\vert C_{YX}^3\rangle$ | $\vert C_{YX}^2\rangle$ | $\vert C_{YX}^1\rangle$ | $\vert C_{YX}^0\rangle$ |
| $\vert 0\rangle$ | $\vert 0\rangle$ | $\vert 0\rangle$ | $\vert c\rangle$ | $\vert c\rangle$ | $\vert b\rangle$ | $\vert 0\rangle$ | $\vert 0\rangle$ | $\vert 0\rangle$ | $\vert 0\rangle$ | $\vert 0\rangle$ | $\vert 0\rangle$ | $\vert a\rangle$ | $\vert a\rangle$ |
| $\vert 0\rangle$ | $\vert 1\rangle$ | $\vert 0\rangle$ | $\vert c\rangle$ | $\vert d\rangle$ | $\vert b\rangle$ | $\vert 0\rangle$ | $\vert 0\rangle$ | $\vert 0\rangle$ | $\vert 0\rangle$ | $\vert 0\rangle$ | $\vert 0\rangle$ | $\vert a\rangle$ | $\vert a\rangle$ |
| $\vert 1\rangle$ | $\vert 0\rangle$ | $\vert 0\rangle$ | $\vert d\rangle$ | $\vert c\rangle$ | $\vert b\rangle$ | $\vert 0\rangle$ | $\vert 0\rangle$ | $\vert 0\rangle$ | $\vert 0\rangle$ | $\vert 0\rangle$ | $\vert 0\rangle$ | $\vert a\rangle$ | $\vert a\rangle$ |
| $\vert 1\rangle$ | $\vert 1\rangle$ | $\vert 0\rangle$ | $\vert d\rangle$ | $\vert d\rangle$ | $\vert b\rangle$ | $\vert 0\rangle$ | $\vert 0\rangle$ | $\vert 0\rangle$ | $\vert 0\rangle$ | $\vert 0\rangle$ | $\vert 0\rangle$ | $\vert a\rangle$ | $\vert a\rangle$ |
| $\vert 0\rangle$ | $\vert 0\rangle$ | $\vert 1\rangle$ | $\vert c\rangle$ | $\vert c\rangle$ | $\vert e\rangle$ | $\vert 0\rangle$ | $\vert 0\rangle$ | $\vert 0\rangle$ | $\vert 0\rangle$ | $\vert 0\rangle$ | $\vert 0\rangle$ | $\vert a\rangle$ | $\vert a\rangle$ |
| $\vert 0\rangle$ | $\vert 1\rangle$ | $\vert 1\rangle$ | $\vert c\rangle$ | $\vert d\rangle$ | $\vert e\rangle$ | $\vert 0\rangle$ | $\vert 0\rangle$ | $\vert 0\rangle$ | $\vert 0\rangle$ | $\vert 0\rangle$ | $\vert 0\rangle$ | $\vert a\rangle$ | $\vert a\rangle$ |
| $\vert 1\rangle$ | $\vert 0\rangle$ | $\vert 1\rangle$ | $\vert d\rangle$ | $\vert c\rangle$ | $\vert e\rangle$ | $\vert 0\rangle$ | $\vert 0\rangle$ | $\vert 0\rangle$ | $\vert 0\rangle$ | $\vert 0\rangle$ | $\vert 0\rangle$ | $\vert a\rangle$ | $\vert a\rangle$ |
| $\vert 1\rangle$ | $\vert 1\rangle$ | $\vert 1\rangle$ | $\vert d\rangle$ | $\vert d\rangle$ | $\vert e\rangle$ | $\vert 0\rangle$ | $\vert 0\rangle$ | $\vert 0\rangle$ | $\vert 0\rangle$ | $\vert 0\rangle$ | $\vert 0\rangle$ | $\vert a\rangle$ | $\vert a\rangle$ |

**Table IX.** Outcomes that the GQIR versions of Figs. 20 or 21 really have according to the corresponding inputs $\vert X_1X_0Y_0\rangle$.

$|d\rangle = [0.6614 \quad -0.25 \quad -0.25 \quad 0.6614]^T$, with

Po|00> = 43.75%, Po|01> = 6.25%, Po|10> = 6.25%, and Po|11> = 43.75%,

$|e\rangle = [0.6959 \quad -0.125 \quad -0.125 \quad 0.6959]^T$, with

Po|00> = 48.4375%, Po|01> = 1.5625%, Po|10> = 1.5625%, and Po|11> = 48.4375%, and $\phi_a$ is trivial in this analysis.

As we can see, none of these outcomes represents a CBS $\{|0\rangle, |1\rangle\}$, as required by the definition of GQIR[26,35,39-42]. Finally, both examples presented in this section show the difference between the positive statements about the benefits of GRQI expressed in Wang's *et al.*[40], and in the Yan & Venegas-Andraca[35] book, versus the resulting evidence of the scientific method, given that, they were not only implemented on Quirk[8] but also on the platforms of IBM Q Experience[3], Rigetti[5], Quantum Programming Studio[6], and Quantum Inspire of QuTech[7], with identical results.

## Multi-channel representation for quantum images (MCQI)

**Theoretical foundations.** In MCQI, the quantum image is presented as[44]:

$$\left|I(n)_{mc}\right\rangle = \frac{1}{2^{n+1}} \sum_{i=0}^{2^{2n-1}} \left|C^i_{RGB\alpha}\right\rangle \otimes |i\rangle, \tag{35}$$

where the color information $\left|C^i_{RGB\alpha}\right\rangle$ that encodes the RGB channels information is defined as:

$$\begin{aligned} \left|C^i_{RGB\alpha}\right\rangle = & \cos\theta^i_R |000\rangle + \cos\theta^i_G |001\rangle + \cos\theta^i_B |010\rangle + \cos\theta^i_\alpha |011\rangle \\ & + \sin\theta^i_R |100\rangle + \sin\theta^i_G |101\rangle + \sin\theta^i_B |110\rangle + \sin\theta^i_\alpha |111\rangle, \end{aligned} \tag{36}$$

where $\{\theta^i_R, \theta^i_G, \theta^i_B\} \in \left[0, \pi/2\right]$ are three angles that encode the colors of the R, G, and B channels of the *i-th* pixel, respectively, and $\theta_\alpha$ is set to zero to make the two coefficients constant ($\cos\theta_\alpha = 1$, and $\sin\theta_\alpha = 0$) to carry no information.

Now, if we compare the equation inside Fig. 5, for the case of FRQI, with Eq.(36), we can make a correct transpilation between both equations without any problem in the implementation of MCQI on IBM Q Experience[3]. In fact, such transpilation results in the exact implementation of MCQI of Fig. 22.

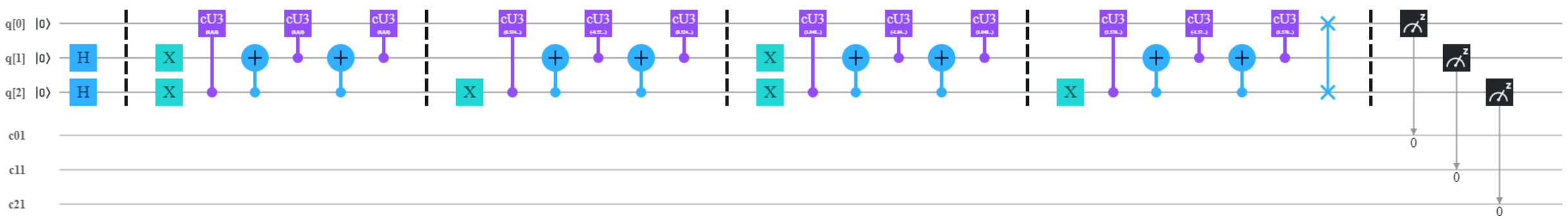

**Figure 22.** Quantum circuit on IBM Q[3] for the transpiled *citiesatnight* code[1] for MCQI.

We resort to the *citiesatnight* code[1] for the correct transpilation from FRQI to MCQI. In fact, this was a very simple procedure since the *citiesatnight* code[1] is in Python, which greatly facilitated the work.

**Implementation on IBM Q[3] Experience.** Implementing the examples of Fig. 7, Table I, and Fig. 8, Table II, corresponding to FRQI, the outcomes are identical on Circuit Composer environment of IBM Q[3]. Therefore, we will use the IBM Q[3] simulator, and one of its 5-qubit quantum processors called

ibmq_burlington, in fairshare Run mode, with 8092 shots. This experiment can be seen in Fig. 23, where, the value of the pixels (0, 255, 0, 255) are equivalent to the angles (0, π/2, 0, π/2). The differences between Fig. 23 (a) and (c) are evident, since, in Fig. 23 (c) there are prohibited places that are occupied {001, 011, 100, 110}, in the same way as in the case of FRQI for the same example. Table X shows the coupling between the pixel values of Table II, their positions (row, column), the angles, and their respective outcomes according to the theoretical outcomes and those of Fig. 24(c). The differences are evident, and once again the statements about the virtues of this technique found in the literature[27, 35, 36, 43-46] are incomprehensible. Finally, Fig. 24 shows the connectivity map of the 5-qubits ibmq_burlington v1.1.4 quantum processing unit (QPU) used in the experiment, which presents a topology with a T form, and whose basis gates are u1, u2, u3, CNOT, and Id.

a)

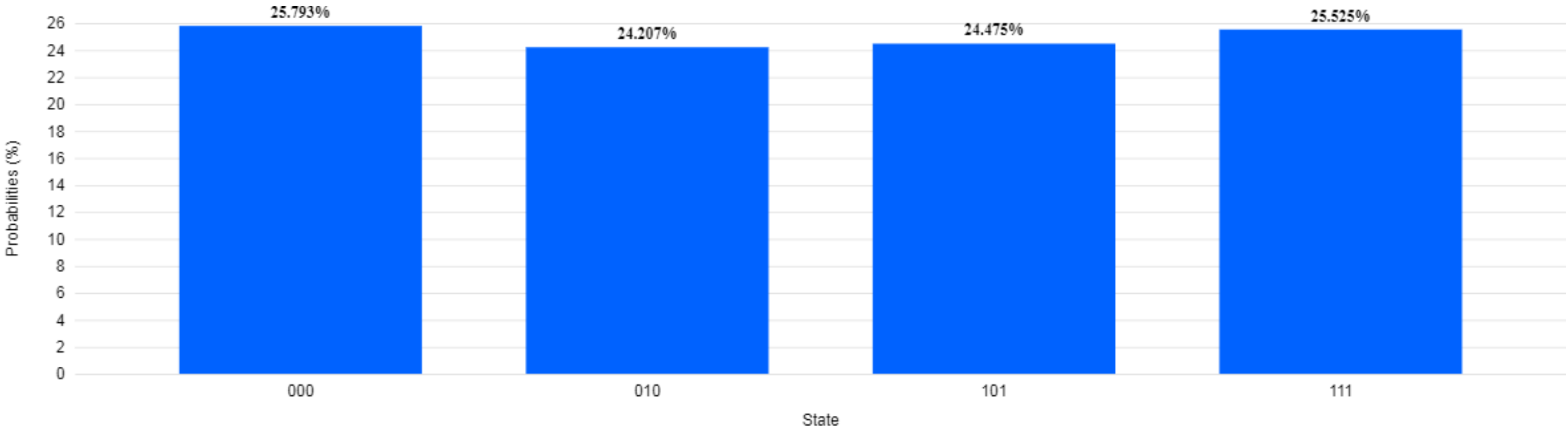

b)

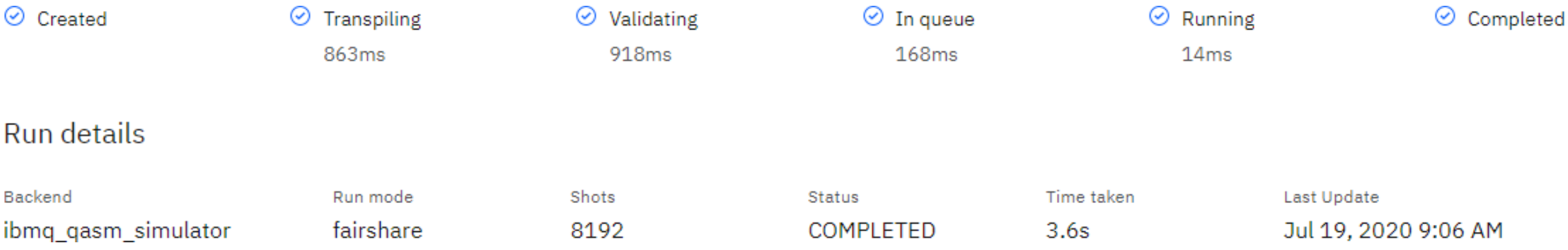

c)

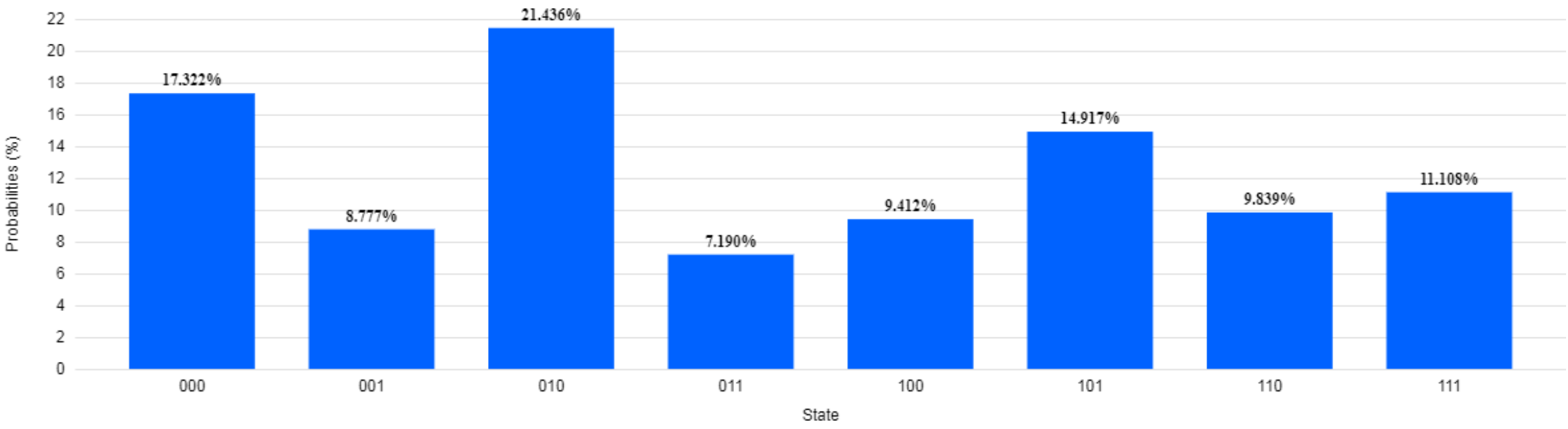

d)

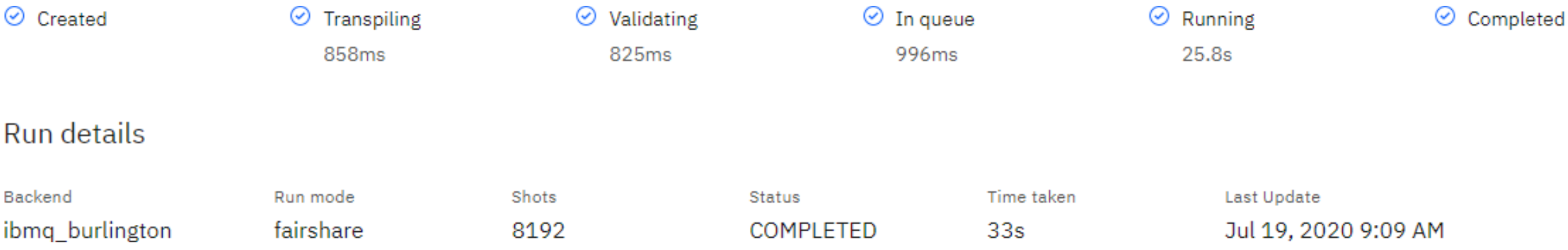

**Figure 23.** Probabilities (outcomes) on IBM Q[3] for the example of Table II, i.e., value of the pixels (0, 255, 0, 255) are equivalent to the angles (0, π/2, 0, π/2): a) on the simulator, b) on the ibmq_burlington (5 qubits) processor, in fairshare Run mode, with 8092 shots.

| row | column | pixels | angles | outcomes (theorists) | outcomes (on the left) | outcomes (on the right) |
|---|---|---|---|---|---|---|
| 0 | 0 | 0 | 0 | 0 | 0.07190 | 0.09412 |
| 0 | 1 | 255 | π/2 | 0.5 | 0.21436 | 0.14917 |
| 1 | 0 | 0 | 0 | 0 | 0.08777 | 0.09839 |
| 1 | 1 | 255 | π/2 | 0.5 | 0.17322 | 0.11108 |

**Table X.** Coupling between the pixel values of Table II, their positions (row, column), the angles, and their respective outcomes according to the theoretical outcomes and those of Fig. 24(c).

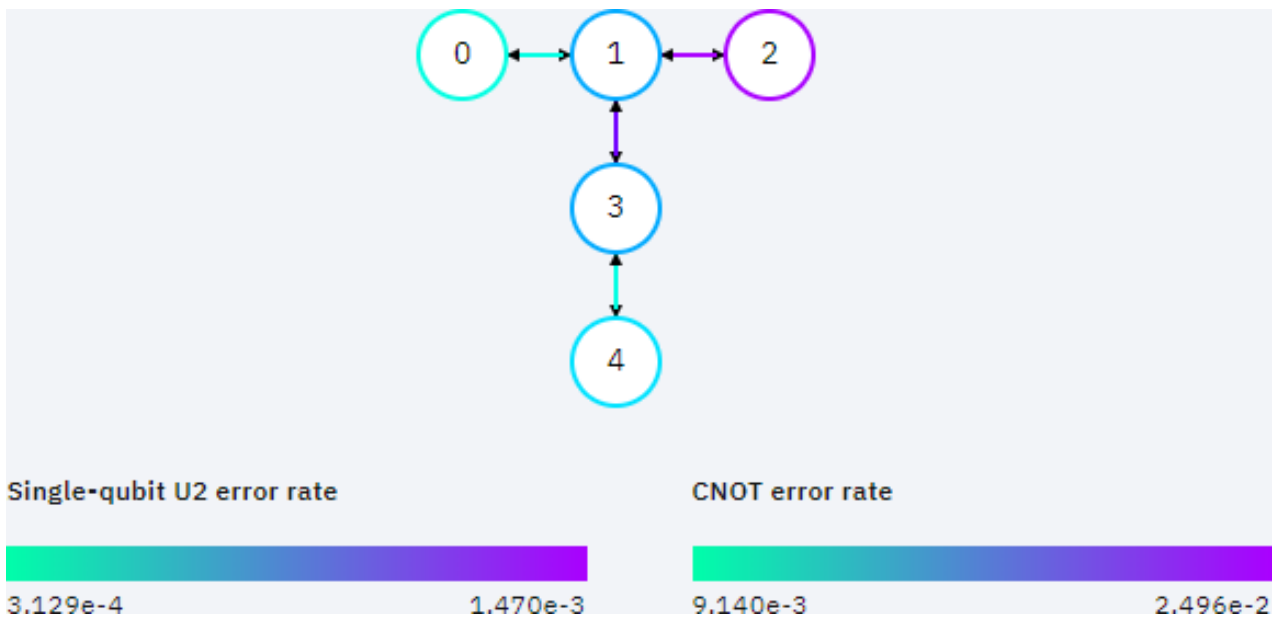

**Figure 24.** Connectivity map of the 5-qubits ibmq_burlington v1.1.4 quantum processing unit (QPU), which includes details of single-qubit U2 and CNOT error rates.

**Quantum Boolean image processing (QBIP)**

**Theoretical foundations.** QBIP[30] establishes a simple and direct transcription between each classic or binary bit $\{0,1\}$, and its corresponding CBS $\{|0\rangle,|1\rangle\}$, without any additional step to the simple execution of a single gate between them: $bit \in \{0,1\} \rightarrow QBIP \rightarrow |bit\rangle \in \{|0\rangle,|1\rangle\}$. A simple way to do this is to have the classic bit become a gate control bit of a Feynman's or CNOT gate[16-18], where the other input of the gate will be an ancilla $|0\rangle$. This representation can be seen in Fig. 25, where a single line represents a quantum wire, while a double line represents a classical wire. Therefore, the complete representation of each pixel of a color image will take place thanks to the use of a complete array of Feynman's gates (CNOT gates) like that of Fig. 25. This will be seen clearly for an example below.

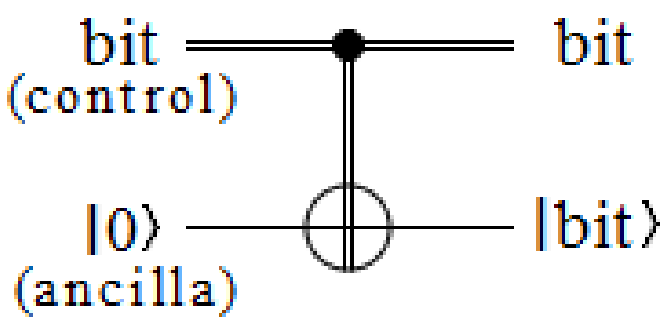

**Figure 25.** QBIP[30] for only one $bit \rightarrow |bit\rangle$ conversion. Classical bit controls the Feynman's gate entering by its upper input, while an ancilla $|0\rangle$ enters by its lower input.

**Implementation on Quirk[8].** Based on the example of Fig. 26 for a hypothetical 2×2 color image with 2 bits per color channel, i.e., with four levels for each color channel, we will implement it in QBIP[30] on Quirk[8] in Fig. 27, in particular, its pixel with green tick with the following set of values: Y = 0, X = 1, R = 11, G = 10, B = 11. We chose Quirk[8] because it is more graphically expressive, however, we had reach similar results with implementations of QBIP[30] on IBM Q Experience[3], Rigetti[5], Quantum Programming Studio[6], and Quantum Inspire by QuTech[7]. Figure 27 shows a perfect coincidence between the two groups of eight metrics to the right of the figure, in order, demonstrating that each bit has its correct counterpart in a perfect CBS, thanks to eight ancillas |0>, and eight CNOT gates. The only limitation is the number of qubits that the QPU supports, a problem common to all techniques.

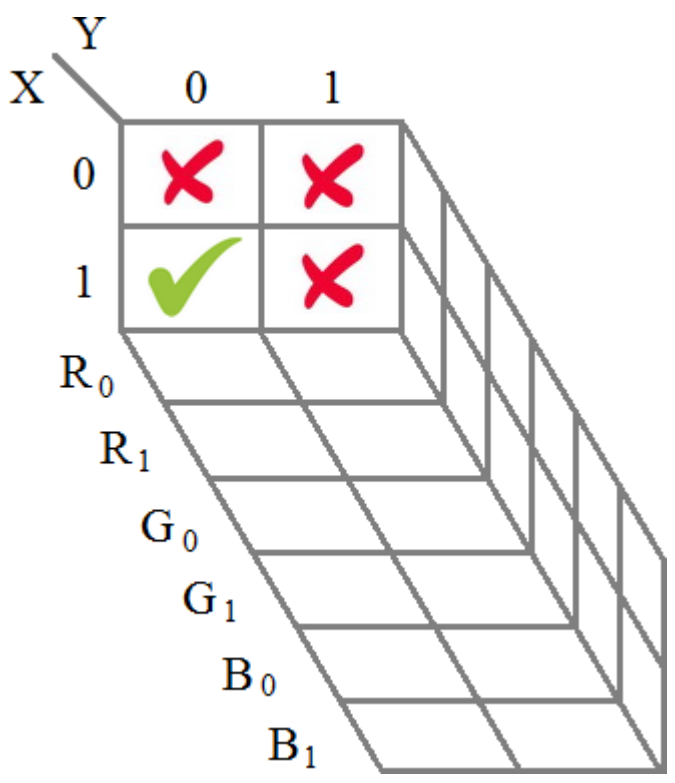

**Figure 26.** Hypothetical 2×2 color image with 2 bits per color channel.

However, as we know from the original QBIP30 paper, in most cases this technique does not require more than the first bit-plane (conformed by the Most Significant Bits: MSB) of each image color channel to perfectly make its job.

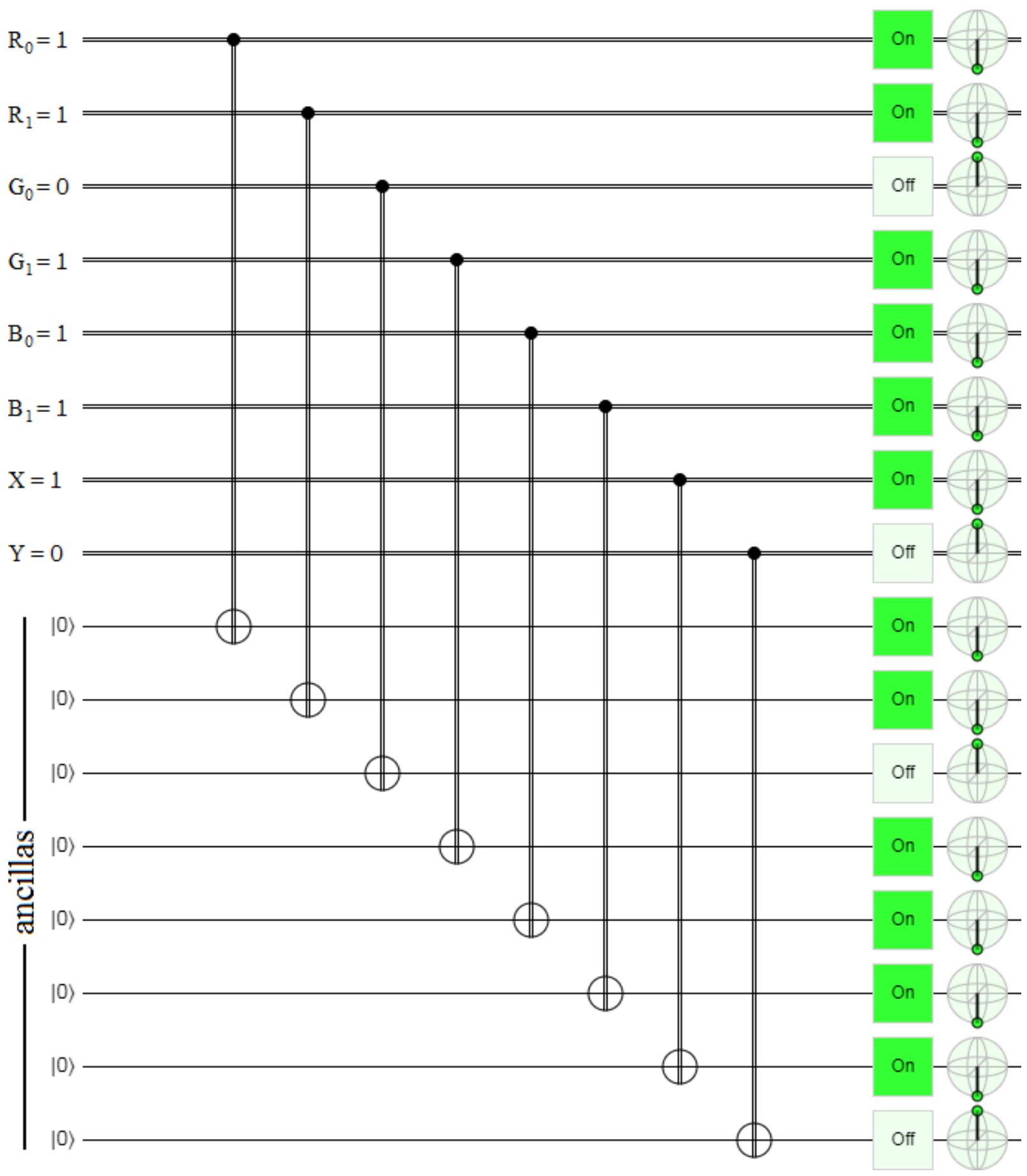

**Figure 27.** Quirk[8] implementation of the example of Fig. 26 for the pixel with the green tick.

## Conclusions

Table III from the Harding and Geetey's version[2] for FRQI[24] implemented on the ibmq_burlington (5 qubits) quantum processor of IBM Q[3] in fairshare Run mode, with 8092 shots; Table V for NEQR[25] implemented on Quirk[8]; Table VII for GQIR[26] from the Wang *et al.* paper[40], and Table IX for GQIR from the Yan & Venegas-Andraca book[35], implemented both on Quirk[8]; and Table X thanks to the transpilation of MCQI[44] from FRQI thanks to the code of *citiesatnight*[1], implemented on the ibmq_burlington (5 qubits) quantum processor of IBM Q[3] in fairshare Run mode, with 8092 shots, they clearly show the problems of wiring techniques versus direct input techniques such as QBIP[30] of Fig. 27. The first conclusions that we can obtain from the wiring techniques is that where NEQR and GQIR are wired, i.e., used, they do not work, in other words, they do not have outcomes of the CBS type $\{|0\rangle,|1\rangle\}$, except in those qubits where both techniques have not been used.

On the other hand, FRQI and MCQI deliver outcomes which are not CBS either and this is because both techniques accumulate too many gates per pixel, with all the inadvisable that this fact represents [29]. Tables III and X show that the accumulated error is scandalously inadmissible in both techniques.

Therefore, in light of the experimental evidence carried out in this work, the claims made in the literature about the virtues of wiring techniques are contradictory[1, 2, 24, 25, 31-40, 44, 47, 50-54].

FRQI and MCQI only work in simulator. This problem was dealt with in a 2017 paper[19] and was not fully understood for the community of Quantum Image Processing. Even more curious is that the same community has created techniques that are supposedly superior to FRQI, such as NEQR, GQIR, and MCQI, where the first two do not even work in the friendliest of simulators, while the third, like FRQI, it works poorly, only on simulators, in an impractical way, against everything accepted and recommended in Quantum Information Processing[29], but at least it does something. They are unintelligible contradictions that take place within the Quantum Image Processing literature[1, 2, 24, 25, 31-40, 44, 47, 50-57], in which what does not work is overexpressed and what works is underexpressed.

Furthermore, it should be mentioned that it is easily verifiable that the problems exposed in this work about wiring techniques extend to all versions, variants and derivatives of FRQI, NEQR, GQIR, and MCQI.

On the other hand, there is not a single Quantum Image Processing paper that evaluates a single quantum error correction technique, which clearly shows a disconnection between the techniques of wiring with practical reality, i.e., their eventual implementations on quantum processing units. Instead, it is extremely easy to carry out a quantum error correction technique[58] on the CNOT gates used in QBIP, which makes it ideal for the exact recovery of hidden data inside images in a steganography[59] context. Finally, QBIP[30] works perfectly on all types of platforms.

**Acknowledgements**
We thank the IBM-Q team for providing us with such simple, convenient and free access to so many tools: circuit composer, simulator, qiskit, as well as so many of their quantum processing units, as those of 1, and 5 qubits, without which this investigation would not have been completed.

**Authors' Contributions**
SSI is responsible for the project's conceptualization, and management, which concludes in this paper. The effort was planned and supervised by SSI and co-supervised by KJKL. SSI ran the research and development team. SSI, KJKL, and MM implement the five techniques on Quirk and on the IBM Q Experience QPU. MM wrote the first version of the paper. SSI, KJKL, and MM analyzed the results. SSI and KJKL reviewed the first version of the paper. SSI, KJKL and MM wrote the final version of the paper. All authors read and approved the final manuscript.

## Competing interests

The authors declare no competing interests.

## Additional information

**Supplementary information** is available for this paper at https://doi.org/10.13140/RG.2.2.25463.65447.

**Correspondence** and requests for materials should be addressed to M.M.